\documentclass[prd,aps,twocolumn,floats,floatfix,nofootinbib]{revtex4-1}

\usepackage{amssymb,bm}
\usepackage{graphicx}
\usepackage{dcolumn}
\usepackage{bm}
\usepackage{graphics}
\usepackage{slashed}
\usepackage{enumerate}
\usepackage{amsmath}
\usepackage{hyperref}
\usepackage{url}
\usepackage[usenames,dvipsnames,svgnames,table]{xcolor}
\usepackage{color}
\usepackage{xcolor}
\usepackage{grffile}

\definecolor{color1}{RGB}{191, 0, 255}

\begin{document}

\title{Turbulent magnetic field amplification in binary neutron star mergers}

\author{
 Carlos Palenzuela$^{1,2,3}$	
 Ricard Aguilera-Miret$^{1,2,3}$,
 Federico Carrasco$^{2,4}$,
 Riccardo Ciolfi$^{5,6}$
 Jay Vijay Kalinani$^{7,6}$
 Wolfgang Kastaun$^{8,9}$
 Borja Mi\~nano$^{1,2,3}$,
 Daniele Vigan\`o$^{10,3,2}$,
 }
 
 \affiliation{${^1}$Departament  de  F\'{\i}sica,  Universitat  de  les  Illes  Balears,  Palma  de  Mallorca, E-07122,  Spain}
  \affiliation{$^2$Institute of Applied Computing \& Community Code (IAC3),  Universitat  de  les  Illes  Balears,  Palma  de  Mallorca, E-07122,  Spain}
  \affiliation{$^3$Institut d'Estudis Espacials de Catalunya (IEEC), 08034 Barcelona, Spain}
  \affiliation{$^4$Instituto de F\'isica Enrique Gaviola, CONICET-UNC, 5000 C\'ordoba, Argentina}
  \affiliation{$^5$INAF, Osservatorio Astronomico di Padova, Vicolo dell'Osservatorio 5, I-35122 Padova, Italy}
  \affiliation{$^6$INFN, Sezione di Padova, Via Francesco Marzolo 8, I-35131 Padova, Italy}
  \affiliation{$^7$Universit\`a di Padova, Dipartimento di Fisica e Astronomia, Via Francesco Marzolo 8, I-35131 Padova, Italy}
  \affiliation{$^8$Max Planck Institute for Gravitational Physics (Albert Einstein Institute), Callinstr. 38, D-30167 Hannover, Germany}
   \affiliation{$^{9}$Leibniz Universit\"at Hannover, D-30167 Hannover, Germany} 
  \affiliation{$^{10}$Institute of Space Sciences (IEEC-CSIC), E-08193 Barcelona, Spain}

\begin{abstract}
Magnetic fields are expected to play a key role in the dynamics and the ejection mechanisms that accompany the merger of two neutron stars. General relativistic magnetohydrodynamic (MHD) simulations offer a unique opportunity to unravel the details of the ongoing physical processes. Nevertheless, current numerical studies are severely limited by the fact that any affordable resolution remains insufficient to fully capture the small-scale dynamo, initially triggered by the Kelvin-Helmholtz instability, and later sourced by several MHD processes involving differential rotation.
Here, we alleviate this limitation by using explicit large-eddy simulations, a technique where the unresolved dynamics occurring at the sub-grid scales (SGS) is modeled by extra terms, which are functions of the resolved fields and their derivatives. The combination of high-order numerical schemes, high resolutions, and the gradient SGS model allow us to capture the small-scale dynamos produced during the binary neutron star mergers, as shown in previous works. Here we follow the first 50 milliseconds after the merger and, for the first time, we find numerical convergence on the magnetic field amplification, in terms of integrated energy and spectral distribution over spatial scales. Among other results, we find that the average intensity of the magnetic field in the remnant saturates at $\sim 10^{16}$~G around $5$~ms after the merger. After $20-30$~ms, both toroidal and poloidal magnetic field components grow continuously, fed by the winding mechanism that provides a slow inverse cascade, i.e. gradually transferring kinetic into magnetic energy. 
We find no clear hints for magneto-rotational instabilities, and no significant impact of the magnetic field on the redistribution of angular momentum in the remnant in our simulations, probably due to the very turbulent and dynamical topology of the magnetic field at all stages, with small-scale components largely dominating over the large-scale ones. Although the magnetic field grows near the rotation axis of the remnant, longer large-eddy simulations are necessary to further investigate the formation of large-scale, helical structures close to the rotational axis, which could be associated to jet formation.
\end{abstract}

\maketitle

\section{Introduction}

The breakthrough binary neutron star (BNS) merger event GW170817 marked the first multimessenger observation of a gravitational wave (GW) source and proved the huge scientific potential of these merging systems \cite{LVC-BNS,LVC-MMA}.
Besides providing the long-awaited compelling evidence that BNS mergers can produce powerful jets and SGRBs~\cite{LVC-GRB,Goldstein2017,Savchenko2017,Troja2017,Margutti2017,Hallinan2017,Alexander2017,Mooley2018a,Lazzati2018,Lyman2018,Alexander2018,Mooley2018b,Ghirlanda2019} as well as a copious amount of heavy r-process elements (e.g.,~\cite{Arcavi2017,Coulter2017,Pian2017,Smartt2017,Kasen2017}; see also~\cite{Metzger2019LRR} and refs.~therein), this event allowed for the first GW-based multimessenger constraints on the neutron star (NS) equation of state (EoS)~\cite{LVC-GRB, LVC-170817properties} and the Hubble constant~\cite{LVC-Hubble}. 
Nonetheless, crucial aspects of the merger and post-merger dynamics remain uncertain, including the actual mechanisms behind jet formation and matter ejection (e.g.,~\cite{Ciolfi2020c}).

Numerical relativity simulations of merging BNSs represent the ideal approach to unveil the details of the physical processes involved in events like GW170817.
In this context, the last decade has been characterized by a rapid progress towards a higher and higher degree of realism with the inclusion of fundamental ingredients such as magnetic fields, temperature and composition dependent EoS, and neutrino radiation (see, e.g.,~\cite{Paschalidis2017,Duez2019,Shibata2019,Ciolfi2020b,Palenzuela2020} for recent reviews). 
In particular, magnetic fields are known to play a key role in the post-merger evolution, in powering relativistic jets, and in shaping the electromagnetic counterpart signals in general (e.g.,~\cite{Ciolfi2020b} and refs.~therein). 
However, magnetohydrodynamics (MHD) increases significantly the complexity of the simulations and poses new computational challenges.

The amplification of magnetic fields during and after merger up to total magnetic energies of the order of $10^{51}$\,erg is due to different mechanisms, including the Kelvin-Helmholtz instability (KHI, e.g.,~\cite{price06,kiuchi15}), the Rayleigh-Taylor instability (RTI, e.g.,~\cite{2021ApJ...921...75S}) and the magneto-rotational instability (MRI,~\cite{balbus91,balbus98,PhysRevLett.96.031101,2013PhRvD..87l1302S,kiuchi14}), that act on very small scales and cannot be fully captured with the current affordable resolutions. Even the highest resolution (and computationally most expensive) simulations being performed nowadays are far from completely resolving these instabilities along with the associated MHD turbulence~\cite{kiuchi18}. Different approaches have been proposed to overcome the above limitations and incorporate at least in part the effects of unresolved MHD processes.

One of the most common approaches is to impose a very strong initial magnetic field $\geq 10^{14}$G (e.g.,~\cite{ruiz16,kiuchi18,ciolfi2019,ciolfi2020collimated,ruiz2020,mosta2020}), many orders of magnitude larger than the values inferred from current observations~\cite{tauris17}, in order to compensate for the inability to capture the KHI-induced amplification. However, the quantitative results obtained within this approach might not be fully reliable, since the amplification via KHI is more efficient on the under-resolved small scales and induces a turbulent flow which should not preserve the initial large-scale ordered field. One promising alternative relies on large-eddy simulations (LESs), a technique in which the evolution equations are modified to account for the unresolved subgrid-scale (SGS) dynamics (e.g.,~\cite{zhiyin15}). Within this approach, the general relativistic MHD (GRMHD) equations are modified by including new terms into the induction equation~\cite{bucciantini13,giacomazzo15,palenzuela15}. Although the results of these studies show an effective growth of the magnetic field, they rely on arbitrarily tuning parameters of the extra terms, losing any predictive character. Other approaches have centered their attention on the turbulent viscous effect induced by the magnetic fields during the post-merger phase, evolving viscous hydrodynamics (HD) in substitution of the MHD equations~\cite{duez2004,shibata2017general,radice17,fujibayashi2020}. These models mimic small-scale dissipation only, therefore they are unable to capture the dynamo mechanism (which usually include additional processes, like kinetic-magnetic energy transfer and magnetic inverse cascade) and depend on physical parameters to be calibrated via very high resolution GRMHD simulations (e.g.,~\cite{radice2020}).

A more sophisticated alternative, based on the so-called gradient SGS model~\cite{leonard75,muller02a}, has been first extended to compressible MHD~\cite{vigano19b}, and more recently also to special and general relativistic MHD~\cite{carrasco19,vigano20}. The gradient SGS model relies on the mathematical Taylor expansion of the non-linear fluxes of the MHD equations as a function of the resolved fields, with no a-priori physical assumptions. The inclusion of these SGS terms in the equations allows to recover, at least partially, the effects that the unresolved sub-grid dynamics induce over the resolved scales. Each SGS term has a free normalization of order unity, which needs to be magnified to compensate for the numerical dissipation of the employed numerical scheme. In our scheme, enhancing such normalization for the induction equation SGS term, allows to improve the convergence of the magnetic amplification excited by the KHI, both in box simulations~\cite{carrasco19,vigano20} and in binary neutron star mergers~\cite{aguilera2020}. 

In this paper, we present a set of simulations with and without SGS terms at different resolutions (finest grid spacing of 120, 60, and 30\,m) for an equal mass BNS merger with a chirp mass that is consistent with GW17018 and a maximum initial magnetic field strength of $5\times10^{11}$\,G. 
The adopted EoS, a hybrid gamma-law EOS based on a piece-wise polytropic approximation of APR4~\cite{read09} (as implemented in~\cite{Endrizzi2016}), leads to a remnant NS which does not collapse to a back hole (BH) within the timespan of the simulations (up to 50\,ms post-merger). 
We study the magnetic field amplification during and after merger and, for the first time, we find convergent results for the magnetic field growth in presence of SGS terms, also consistent with the highest resolution result without SGS terms.
We then investigate the impact of magnetic fields on various aspects of the post-merger dynamics, including the evolution of the differential rotation profile of the remnant NS, the prospects for jet formation, matter ejection, and post-merger GW emission.

The paper is organized as follows. Section \ref{sec:setup} introduces our GRMHD LES approach, the numerical methods and setup, the initial data, and a number of useful analysis quantities. In Section \ref{sec:picture}, we recall a few fundamental notions that are at the base of our results, which are then presented in detail in Section \ref{sec:results}. Finally, our concluding discussion is given in Section \ref{sec:discussion}.

\section{Numerical setup}\label{sec:setup}

\subsection{Evolution equations: GRMHD LES} \label{sec:equations}

The concept and the mathematical foundations behind LES with a gradient SGS approach have been extensively explained in our previous works (and references therein) in the context of Newtonian~\cite{vigano19b} and relativistic MHD \cite{carrasco19,vigano20,aguilera2020}, to which we refer for details and further references. In the present work, we will evolve the full GRMHD equations with Gradient SGS terms, exactly as described in our previous paper \cite{aguilera2020}
(but using a different EoS that will be introduced in Sec.~\ref{sec:EoS}).

The spacetime geometry is described by the Einstein equations.
The covariant field equations can be written as evolution equations by performing the $3+1$ decomposition (see, e.g., \cite{bonabook}),
in which the line element can be written as
\begin{equation}
ds^2 = - \alpha^2 \, dt^2 + \gamma_{ij} \bigl( dx^i + \beta^i dt \bigr) \bigl( dx^j + \beta^j dt \bigr)~, 
\label{3+1decom}  
\end{equation}
where $\alpha$ is the lapse function, $\beta^{i}$ the shift vector, $\gamma_{ij}$ the induced 3-metric on each spatial slice, and $\sqrt{\gamma}$ is the square root of its determinant.
In this work, we use the covariant conformal Z4 formulation of the evolution equations \cite{alic12,bezares17}.
A summary of the final set of evolution equations for the spacetime fields, together with the gauge conditions setting the choice of coordinates, can be found in \cite{palenzuela18}.

The GRMHD equations for a magnetized, non-viscous and perfectly conducting fluid~\cite{palenzuela15} 
provide a set of evolution equations for the conserved variables $ \left\lbrace \sqrt{\gamma}D, \sqrt{\gamma}S^i , \sqrt{\gamma}U, \sqrt{\gamma}B^i \right\rbrace$. These conserved fields are functions of the rest-mass density $\rho$, the specific internal energy $\epsilon$, the velocity vector $v^{i}$ and the magnetic field $B^i$ (primitive fields), namely:
\begin{eqnarray}
D &=& \rho W ~, \\
S^i &=& (h W^2 + B^2 ) v^i - (B^k v_k) B^i ~,\\
U &=& h W^2 - p + B^2 - \frac{1}{2} \left[ (B^k v_k)^2 + \frac{B^2}{W^2} \right]  ~,
\end{eqnarray}
where $W = (1-v^2)^{-1/2}$ is the Lorentz factor. The pressure $p$ is obtained from the EoS as detailed in Sec.~\ref{sec:EoS}. 

The full set of evolution equations, including the hyperbolic divergence cleaning via damping of a new field $\phi$~\citep{palenzuela18} and all the gradient SGS terms, can be found in~\cite{vigano20,aguilera2020}. Following our previous studies, and since we are mostly interested in the magnetic field dynamics, we include only the SGS term appearing on the induction equation,
\begin{eqnarray}
&& \partial_t (\sqrt{\gamma} {B}^i) + \partial_k [\sqrt{\gamma}(- \beta^k {{B}}^i  +  \beta^i {{B}}^k) 
\nonumber \\
&& \quad\quad\quad\quad + \alpha \sqrt{\gamma} ({\gamma}^{ki} {{\phi}} + {M}^{ki} - {\tau}^{ki}_{M} )] = \sqrt{\gamma} {{R_B}}^i ~,
\label{evol_B_sgs}
\end{eqnarray}
where $M^{ki} = B^{i} v^{k} - B^{k} v^{i}$, while the source term ${{R_B}}^i$ is related to the divergence cleaning field (see \cite{vigano20} for its explicit expression). The SGS tensor is given by 
\begin{equation}
\tau^{ki}_{M} = -~{\cal C_M}~\xi \, H_{M}^{ki} ~~, \label{eq:sgs_gradient}
\end{equation}
where the explicit expressions for the tensor $H_{M}^{ki}$ in terms of field gradients are quite lengthy and can be found in the Appendix~A of \cite{aguilera2020}.

Although the pre-coefficient ${\cal C_M}$ is meant to be of order one for a numerical scheme having a mathematically ideal Gaussian filter kernel and neglecting higher-order corrections, the value that best mimics the feedback of small scales onto the large scales in a LES can differ depending partially on the numerical methods employed and on the specific problem, as discussed in \cite{vigano19b,carrasco19,vigano20, aguilera2020}. In this work, we will set ${\cal C_M}= 8$, which has been shown to reproduce the magnetic field amplification more accurately (i.e., comparing with very high-resolution simulations) for our numerical schemes~\cite{vigano20,aguilera2020}. Note that less dissipative numerical schemes would need smaller values of ${\cal C_M}$.
Hereafter, we refer to this particular choice of SGS terms as LES, as opposed to a standard simulation with no SGS terms (i.e., ${\cal C_M}= 0$). The coefficient $\xi= \gamma^{1/3} \Delta^2/24$ is proportional to the spatial grid-spacing squared --typical for SGS models-- and hence ensures by construction the convergence to the continuous limit. 

\subsection{Numerical methods}

As in our previous works, we use the code {\sc MHDuet}, generated by the platform {\sc Simflowny} \cite{arbona13,arbona18} and based on the {\sc SAMRAI} infrastructure \cite{hornung02,gunney16}, which provides the parallelization and the adaptive mesh refinement. The code has been deeply tested for different scenarios \cite{palenzuela18,vigano19,vigano20,liebling20}, including basic tests of MHD and GR. In short, it uses: fourth-order-accurate operators for the spatial derivatives in the SGS terms and in the Einstein equations (the latter are supplemented with sixth-order Kreiss-Oliger dissipation); a high-resolution shock-capturing (HRSC) method for the fluid, based on the  Lax-Friedrich flux splitting formula \cite{shu98} and the fifth-order reconstruction method MP5 \cite{suresh97}; a fourth-order Runge-Kutta scheme with sufficiently small time step $\Delta t \leq 0.4~\Delta$ (where $\Delta$ is the grid spacing); and an efficient and accurate treatment of the refinement boundaries when sub-cycling in time~\cite{McCorquodale:2011,Mongwane:2015}.  A complete assessment of the implemented numerical methods can be found in \cite{palenzuela18,vigano19}.

The binary is evolved in a cubic domain of size $\left[-1204,1204\right]$~km. The inspiral is fully covered by $7$ Fixed Mesh Refinement (FMR) levels. Each consists of a cube with twice the resolution of the next larger one. In addition, we use up to $2$ Adaptive Mesh Refinement (AMR) levels,  covering the regions where the density is above $5 \times 10^{12}~\rm{g~cm^{-3}}$ and providing a uniform resolution throughout the shear layer. We compare simulations with 0--2 AMR levels, labeled LR,MR,HR, but same FMR grid. This set up allows us to study the convergence in the bulk region, which is our main interest, while saving resources in the outer envelope.
With this grid structure, we achieve a maximum resolution of $\Delta_{min}$ in a domain $\Delta L_{min}$ covering at least the most dense region of the remnant.
The specific values of these grid parameters, for the different resolutions considered here, can be found in Table~\ref{tab:models}. 

\subsection{EoS and conversion to primitive variables} \label{sec:EoS}

We consider a hybrid EoS during the evolution, with two contributions to the pressure. For the cold part, we use a tabulated version of the piecewise polytrope fit to the APR4 zero-temperature EoS~\cite{read09}, with a modification to prevent superluminal speeds~\cite{Endrizzi2016}.
The contributions of thermal effects are modeled by the gamma-law $p_{th} = (\Gamma_{th} - 1) \rho \epsilon$, with adiabatic index $\Gamma_{\rm th}= 1.8$.

The conversion from the evolved or conserved fields to the primitive or physical ones is performed by using the  robust procedure introduced in \cite{kastaun20}. Following a common practice in GRMHD simulations, the surrounding regions of the neutron stars are filled with a relatively tenuous, low-density {\em atmosphere}, which is necessary to prevent the failure of the HRSC schemes usually employed to solve the MHD equations. To minimize unphysical states of the conserved variables outside the dense regions, produced by the numerical discretization errors of the evolved conserved fields, we enforce a minimum density in the atmosphere of $6 \times 10^{5}~\rm{g~cm^{-3}}$, while setting the velocity to zero and the magnetic field to its previous value in these regions.

In addition, we apply the SGS terms only in regions where the density is higher than $6 \times 10^{13}~\rm{g~cm^{-3}}$ in order to avoid possible spurious effects near the stellar surface and in the atmosphere. Since the remnant's maximum density is above $10^{15} ~\rm{g~cm^{-3}}$, the SGS model is still applied to a considerable volume, but only in the most dense regions of the star. Even though, it still allows us to capture the MHD instabilities leading to a strong amplification of the magnetic field soon after the merger. In any case, we would like to remind that our grid set up allows us to study the convergence only in the bulk region.

\subsection{Initial conditions}

The initial data is created with the {\sc Lorene} package~\cite{lorene}, using the APR4 zero-temperature EoS described above. We consider an equal-mass BNS in quasi-circular orbit, with an irrotational configuration, a separation of $45$ km and an angular frequency of $1775\ \rm{rad~s^{-1}}$. The chirp mass $M_{chirp}=1.186~M_{\odot}$ is the one inferred in the refined analysis of GW170817 \cite{LVC-170817properties}, implying a total mass $M=2.724~M_{\odot}$ for the equal mass case. 

Each star initially has a purely poloidal dipolar magnetic field that is confined to its interior, calculated from a vector potential 
$A_ {\phi} \propto R^2 (P - P_{cut})$, where $P_{cut}$ is a hundred times the pressure of the atmosphere, and $R$ is the distance to the axis perpendicular to the orbital plane passing through the center of each star. The maximum magnetic field (at the centers) is $5\times 10^{11}$ G. This is orders of magnitude lower than the large initial fields used in other simulations (e.g.,~\cite{kiuchi15,ruiz16,kiuchi18,ciolfi2019,ciolfi2020collimated,ruiz2020}) and not too far from the upper range of expectations for old NSs (e.g.,~\cite{2014ApJS..212....6O}). Such values are also at the lower border of computational feasibility, since the accurate evolution for too small ratios of magnetic-to-kinetic pressure is hampered by discretization errors.
Nevertheless, the initial topology is quickly forgotten after merging: acting only as a seed for the KHI, it has a negligible effect on the final magnetic field configuration of the remnant, as long as the initial values are not too large ($B\lesssim 10^{13}$ G for a medium resolution), as we show in an accompanying paper~\cite{aguilera21}.

\subsection{Analysis quantities}

Below, we introduce definitions of several integral quantities that we use to monitor the dynamics in different regions. For instance, we compute suitable averages of the magnetic field strength; the fluid angular velocity $\Omega \equiv \frac{d\phi}{dt} = \frac{u^{\phi}}{u^t}$ (where $u_a \equiv W (-\alpha, v_i )$
denotes the fluid four-velocity); and the plasma beta parameter, $\beta = \frac{2 P}{B^2}$. The averages for a given quantity $q$ over a certain region ${\cal N}$ 
will be denoted generically by, 
\begin{eqnarray}
 {<}q{>}_a^b = \frac{\int_{{\cal N}} q \, d{\cal N}}{\int_{{\cal N}} d{\cal N}} ,
\end{eqnarray}
where ${\cal N}$ stands for a volume $V$, a surface $S$, or a line $\ell$, 
and the integration is restricted to regions where the mass density is within the range $(10^a, 10^b)\, {\rm g/cm^3}$. If $b$ is omitted, it means no upper density cut is applied.
In particular, we define averages over the bulk of the remnant as ${<}q{>}_{13}$, and averages over the envelope as ${<}q{>}^{13}_{10}$. Surface integrals are carried out over a cylinder $S$ with axis passing through the center of mass\footnote{The center of mass is calculated using the non-relativistic formula.} and orthogonal to the orbital plane. Line integrals are carried out along circles $\ell$ around the same axis. In cylindrical coordinates $(R,\phi,z)$, the relevant surface and line elements are, respectively:
\begin{eqnarray*}
dS &=& \sqrt{\gamma_{zz}\gamma_{\phi\phi} - \gamma_{z\phi}^2} \, d\phi dz ,\\ 
d\ell &=& \sqrt{\gamma_{\phi \phi}} \, d\phi .
\end{eqnarray*}

We also compute global quantities, integrated over the whole computational domain,
such as the total baryonic mass $E_{bar}$, magnetic energy $E_{mag}$, thermal energy $E_{th}$ and rotational kinetic energy $E_{rot}$, defined by:
\begin{eqnarray}
E_{bar} &=& \int \rho W   \sqrt{\gamma} dx^3 ,\\
E_{mag} &=& \frac{1}{2}\int B^2 \sqrt{\gamma} dx^3 ,    
\\
E_{th} &=& \int \rho W (\epsilon - \epsilon_{cold}) \sqrt{\gamma} dx^3 ,
\\
E_{rot} &=& \frac{1}{2}\int \Omega T^t_{\phi} \sqrt{\gamma} dx^3 ,
\end{eqnarray}
where $T^t_{\phi}$ are just the time-azimuthal components of the stress-energy tensor for a perfect fluid.

An informative way to analyze the results is by computing the distribution of the kinetic and magnetic energies over the spatial scales, simply called spectra hereafter, and defined respectively as:
\begin{eqnarray}
&& \mathcal{E}_k(k) = \frac{L^3 4\pi}{(2\pi)^3N^6} <k^2|\widehat{\sqrt{\rho}\vec{v}}|^2(\vec{k})>_{k}~, \label{eq:spectra_k}\\
&& \mathcal{E}_m(k) = \frac{L^3 4\pi }{(2\pi)^3N^6}<k^2|\hat{\vec{B}}|^2(\vec{k})>_{k}~, ~\label{eq:spectra_b}
\end{eqnarray}
where $L=32$ km is the size of a dominion containing the densest part of the remnant over which we perform the fast Fourier transform (indicated here by the wide hat), $k \in [2\pi/L,\pi/\Delta]$ is the radial wavenumber, $<>_k$ is the average over the spherical surface in the Fourier space corresponding to a fixed radial wavenumber $k_r = k$, and $N = L/\Delta$ is the number of points per each direction (ensuring the correct normalization). For the magnetic spectra, we also calculate the poloidal and toroidal contributions separately. Further details of the numerical procedure to calculate the spectra can be found in~\cite{aguilera2020,vigano19,vigano20}.
Notice that in those works the spectra was  calculated using only the solution within the bulk of the remnant (i.e., by setting to zero the fields below a certain threshold density). Here, in order to preserve the large wavelength modes and to include at least partially the solution in the outer envelope, we decided not to modify the fields whatsoever. This choice comes at the price of not having periodic boundary conditions in our dominion, which will induce a artificial pile-up of energy at very high wave-numbers.
With these spectrum distributions we can define the spectra-weighted average wavenumber
\begin{equation}
  {\bar k} \equiv \frac{\int_k k\,{\cal E}(k) \,dk} {\int_k {\cal E}(k)\, dk}
\end{equation}
with an associated length scale $\delta R = 2\pi/{\bar k}$ which represents the typical coherent scale of the structures present in the field.

Although it is not the main goal of this article, we also compute the ejected mass and the GWs emitted by the system. The ejected mass is  estimated from the flux of unbound mass crossing a sphere of radius $r$ far away from the source, namely
\begin{equation}
{\dot M}_{ej}=\int\alpha J^r r^2\sqrt{\gamma} d\Omega  ~~,~~
J^r = \rho W \left( v^r-\frac{\beta^r}{\alpha} \right)
\end{equation}
The gravitational radiation is described in terms of the Newman-Penrose scalar $\Psi_{4}$, which can be expanded in terms of spin-weighted $s=-2$ spherical harmonics~\cite{rezbish,brugman}, namely 
\begin{equation}
r \Psi_4 (t,r,\theta,\phi) = \sum_{l,m} C_{l,m}(t,r) \, Y^{-2}_{l,m} (\theta,\phi).
\label{eq:psi4}
\end{equation}
The coefficients $C_{l,m}$ are extracted from spherical surfaces at different extraction radii $r_{ext}=(150,300,450)$~km.
The angular frequency of each gravitational-wave mode can be calculated easily from: 
\begin{eqnarray}
	\omega_{l,m} = - \Im \left(\frac{{\dot C}_{l,m}}{C_{l,m}}\right).
\end{eqnarray}
We recall that the angular frequencies $\omega_\mathrm{src}$ of remnant oscillations are related to
the resulting GW components by $\omega_{l,m} = m \omega_\mathrm{src}$. 

\section{The big picture}\label{sec:picture}

The dynamics during the merger and post-merger is very rich, so one could expect several MHD processes operating to amplify and sustain the magnetic field in timescales up to hundreds of milliseconds. These processes have been analytically studied under idealized conditions: however, since our scenario is very non-linear and highly dynamical, it is not easy to clearly distinguish them. First, the shear layer that forms when the surfaces of the NSs touch is subject to the KHI. The Rayleigh-Taylor instability (RTI) could also play a role and together they can generate a turbulent state, that can lead to a small-scale dynamo. Secondly, the MRI~\cite{balbus91,balbus98} can also amplify the magnetic field and maintain a turbulent state in the remnant, as suggested by global numerical simulations~\cite{PhysRevLett.96.031101,2013PhRvD..87l1302S,kiuchi14}. In addition, the differential rotation tends to wind up the magnetic field lines. Here we review some of the relevant aspects.

\begin{itemize}
	\item Kelvin-Helmholtz instability (or dynamical shear instability): in a shear layer of thickness $d$ between two fluids moving with opposite directions, all the modes with wavelength $\lambda > d$ are going to be unstable and produce vortexes with a growth rate given by~(see, e.g.,~\cite{price06})
	\begin{equation}
	\sigma = \frac{\pi \Delta v}{\lambda}
	\end{equation}
	where $\Delta v$ is the jump in velocity across the shear layer. As a result of this vortex motion, magnetic fields are wound and their strength are amplified in an exponential manner within a timescale given by 	
	\begin{equation}
	t_\mathrm{KHI} \sim \sigma^{-1} \sim 10^{-3} \,\mathrm{ms} \left( \frac{\lambda}{1\,\mathrm{km}}\right) \left( \frac{0.5 c}{\Delta v}\right)
	\end{equation}
	which is much smaller than the characteristic timescale of the system. Note that smaller wavelengths grow faster, inducing a turbulent state that further amplifies the magnetic energy via small-scale dynamo processes. 
    \item Rayleigh-Taylor instability: it operates on the interface between two fluids where the density gradient is misaligned with
	the direction of the local gravitational field. The instability characteristically evolves with rising bubbles of less dense fluid and sinking spikes of more dense fluid that propagate away from the interface. The nonlinear interaction of the bubbles and the spikes induces a region of turbulent mixing. It has been recently proposed~\cite{2021ApJ...921...75S} that the small-scale dynamo driven by RTI turbulence might help to explain the observed magnetic energy amplification, on the time scale of microseconds, developing in the outer regions
	of the remnant, which would act complementary to  amplification by the KHI observed mostly in the core. 	
	
	\item Magneto-rotational instability (MRI, or magnetic shear instability): originally studied for disks \cite{balbus91}, it requires a differential rotation, $\partial_R \Omega < 0$, e.g. in a Keplerian disk, $\Omega = \Omega_K \sim R^{-3/2}$.
	Assuming idealized condition of a very smooth large-scale magnetic field (usually uniform in the vertical or azimuthal direction, or dipolar), some modes can quickly grow at the expenses of the rotational energy. These modes consist of adjacent channels having different angular momentum and magnetic fluxes, with a spatial width corresponding to the fastest growing MRI modes. At the end of the first exponential growth phase, the instability saturates and the channels dissolve into a turbulent flow featuring a complex magnetic field topology. These channel-flow structures (and the subsequent turbulent state, which increases the effective viscosity of the fluid) redistribute angular momentum from internal regions to external ones. The characteristic timescale and wavelength for the fastest growing mode may be estimated (see, e.g.,~\cite{shibatabook}) by
	\begin{align}
	\begin{split}
	  t_\mathrm{MRI} &= 2 \Big| \frac{\partial \Omega}{\partial \ln R} \Big|^{-1} \approx \frac{4}{ 3 \Omega} \\
	  & \approx  0.1 \,\mathrm{ms} \left( \frac{10\, \mathrm{rad}/\mathrm{ms}}{\Omega}\right)
	\end{split}\\
	\begin{split}	
	\lambda^i_\mathrm{MRI} &\approx 
	\frac{2 \pi}{\Omega} v^i_A \approx 
	  600 \,\mathrm{m} \left( \frac{10\, \mathrm{rad}/\mathrm{ms}}{\Omega}\right) \\
	  &\qquad\times \left( \frac{{\bar B}^i}{10^{16}\,\mathrm{G}}\right) \left( \frac{10^{15} \,\mathrm{g}/\mathrm{cm}^{3}}{{\bar \rho}} \right)^\frac{1}{2} 
	\end{split}
	\end{align}
	where $v^i_A = {\bar B}^i/\sqrt{4 \pi {\bar \rho}}$ is the Alfven speed in the direction $x^i$ and we have assumed a Keplerian profile. Here ${\bar B}^i,{\bar \rho}$ represent the mean values (i.e., in space and time) of the magnetic field and the density. Note that in a turbulent flow  ${\bar B}^i=0$ and the previous estimates are meaningless for two reasons. The first one is that it is necessary to introduce and effective Alfven speed, as we will discuss later. The second one is that the fundamental assumption of the linear analysis leading to the instability phase requires idealized, very smooth (i.e., over lengthscales much larger than $\lambda_{\rm MRI}$) and static (i.e., over timescales much longer than $t_{\rm MRI}$) background magnetic fields. As we will show, this condition is very hardly satisfied in the BNS post-merger scenario, consisting since the very beginning of a very dynamical, small-scale dominated magnetic field. Although the MRI could still play a role, the quantitative estimations of the fast growing modes $\lambda_{MRI}$ are in general not reliable.
 	
	\item Winding up effect: for axisymmetric configurations it is easy to show (see, e.g.,~\cite{shibatabook}), using the induction equation and the solenoidal constraint, that the toroidal magnetic field $B^{T} \equiv R B^{\phi}$ will grow linearly as
	\begin{eqnarray}\label{B_winding}
	|B^T| = t R \Big| B^R  \frac{\partial \Omega}{\partial R} \Big|
	\approx \frac{3}{2} \Omega t |B^R|
	\end{eqnarray}
	where we have assumed again a Keplerian profile to obtain the last relation. The twisting of the magnetic fields increases its tension, which will grow until it can eventually brake the fluid motion, approximately when the toroidal magnetic energy $B_T^2/(4 \pi)$ reaches the rotational energy $\rho R_*^2 \Omega^2$ ($R_*$ being the typical radius of the remnant). Using the expression Eq. (\ref{B_winding}) for $B^T$, the magnetic braking occurs in a timescale given by
	\begin{multline}
	t_\mathrm{brake} \sim \frac{R_*}{v^R_A} \approx 
	 15 \,\mathrm{ms}  \left( \frac{R_*}{15 \,\mathrm{km}}\right) \\
	 \times \left( \frac{10^{16} \,\mathrm{G}}{B}\right) \left( \frac{\rho}{10^{15} \,\mathrm{g} /\mathrm{cm}^{3}} \right)^\frac{1}{2} 
	\end{multline}
	
\end{itemize}

\begin{figure*}
	\centering
	\includegraphics[width=0.32\textwidth]{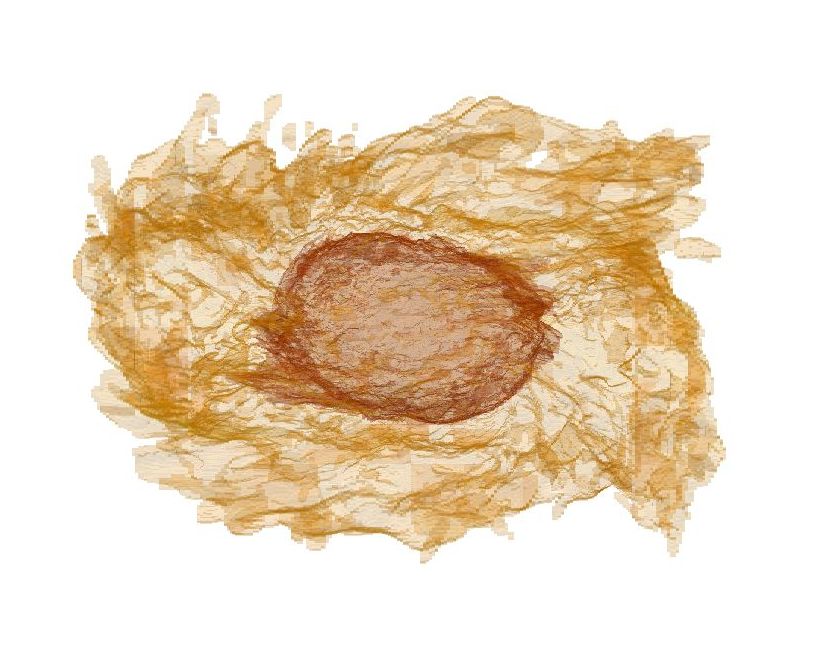}
	\includegraphics[width=0.32\textwidth]{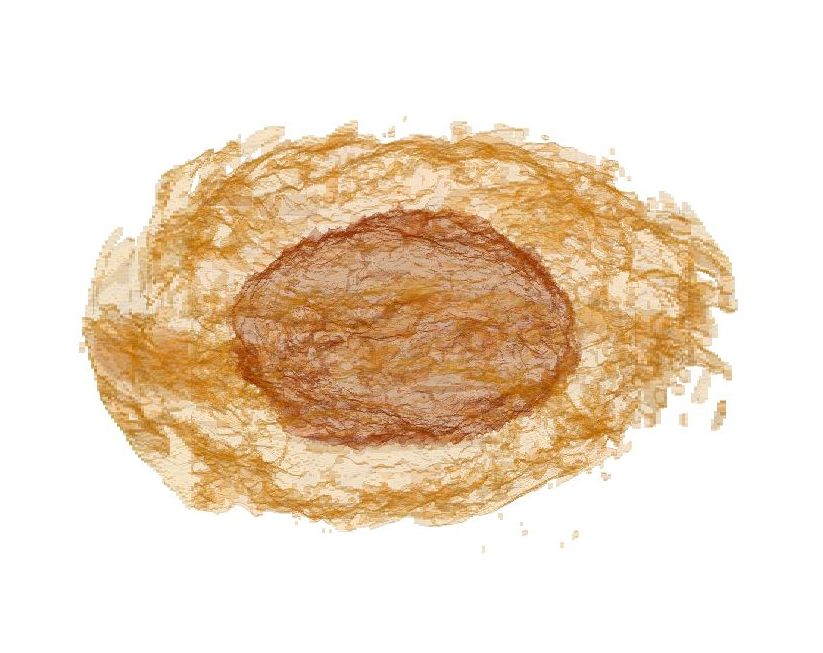} 
	\includegraphics[width=0.32\textwidth]{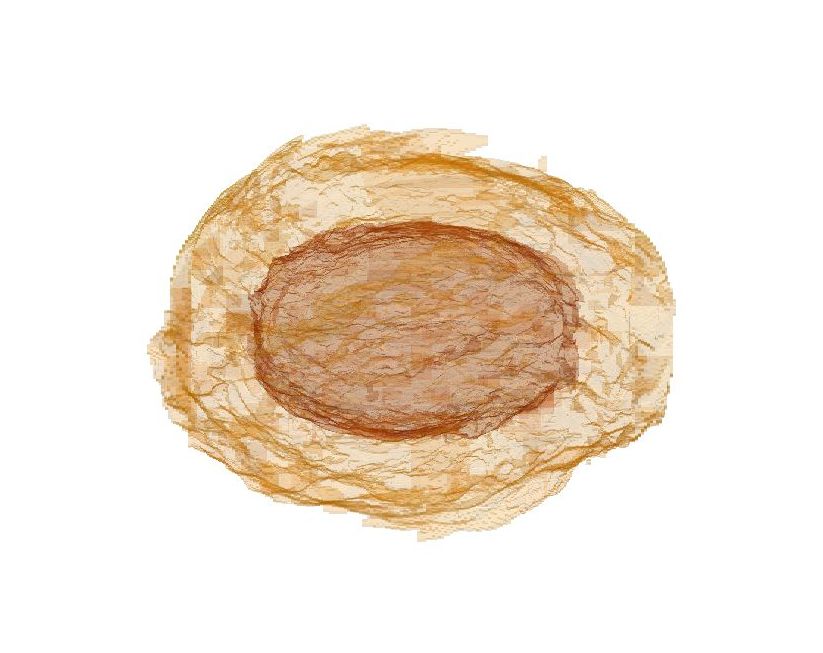} 	
	\caption{ {\em Surfaces of constant density in the remnant}. From left to right, we show representative times $t=(4,6,9)$~ms after the merger. The collision produces a rotating massive remnant that eventually settles down into a spinning neutron star surrounded by an extended torus or outer envelope. The inner surface, at $\rho=10^{13} \mathrm{g}/\mathrm{cm}^{3}$, shows the transition between the bulk and envelope. The outer surface, at $\rho=10^{12} \mathrm{g}/\mathrm{cm}^{3}$, displays (partially) the shape of the envelope.}
	\label{fig:density3D}
\end{figure*}	

Note that most of the previous estimates are performed under the linear analysis of a background smooth, idealized magnetic field topology (e.g., purely vertical, or toroidal) and fluid flow. This does not necessarily apply to the situation after merger. For example, the vorticity field of the disk studied in~\cite{kastaun21} is quite irregular. The MRI and the winding estimates use the Alfven speed, which implies a mean magnetic field ${\bar B}^i$ coherent in spatial and time scales much larger than those of the instabilities.
In turbulent flows, the magnetic field is small scale, randomly oriented, meaning that ${\bar B}^i=0$. In such flows, the waves can not propagate, at least during a relatively long timescale, in any specific direction. Nevertheless, one can still define an effective Alfven velocity. For instance, one can assume that the linear analysis is valid in a size given by the length of the characteristic vortices of the turbulent magnetic-field~\cite{kiuchi18}. A more restrictive condition is to consider that the perturbations
undertake a random walk. In both cases, the effective Alfven velocity can be written in the same generic way, 
\begin{eqnarray}
v^i_A = \frac{{\bar B}^i}{\sqrt{4 \pi {\bar \rho}}} \left( \frac{\delta R}{R_*} \right)^{n}
\end{eqnarray}
with $n=1$ for the mean values valid in the  characteristic spatial scale of the turbulent magnetic field $\delta R$ and $n=2$ for the random walk estimate. By considering values $\delta R \approx 750$ m, we obtain that $v_A$ is modified by a factor $0.05-0.0025$, depending on the value of $n$. 
This means that the modified $t_\mathrm{brake}$ is extended by one or two orders of magnitude, indicating that the winding will fully develop in much longer timescales of hundreds of milliseconds. Along the same lines, it would imply also that the $\lambda^i_\mathrm{MRI}$ is decreased by the same factor, making it even more challenging to resolve it. Even in that case, $t_\mathrm{MRI}$ does not depend on the Alfven velocity. Therefore, if the field moves in shorter timescales, the MRI can not fully develop. These considerations imply that the MRI could act only if the field is not too turbulent and is well organized at large scales.
We will further discuss these issues by comparing with the results of our simulations in the conclusions.

\section{Results}\label{sec:results}

\begin{table}[t]
	\begin{tabular}{ |c|c|c|c|c| }			
		\hline
		Case
		& ${\cal C_M}$
		& Refinement levels
		& $\Delta L_{min}$ [km]
		& $\Delta_{min}$ [m]
		\\ \hline
		{\tt LR} & 0 & 7 FMR & [-28,28] & 120 \\
		{\tt MR} & 0 & 7 FMR + 1 AMR & [-13,13] & 60 \\
		{\tt HR} & 0 & 7 FMR + 2 AMR & [-11,11] & 30 \\
		{\tt LR LES} & 8 & 7 FMR & [-28,28] & 120 \\
		{\tt MR LES} & 8 & 7 FMR + 1 AMR & [-13,13] & 60 \\
		{\tt HR LES} & 8 & 7 FMR + 2 AMR & [-11,11] & 30 \\
		{\tt MR B0} & 8 & 7 FMR + 1 AMR & [-13,13] & 60 \\
		\hline
	\end{tabular}
	\caption{{\em Parameters of the simulations:} different resolutions, mesh refinement setup (with the finest grid spacing $\Delta_{min}$ covering a region of size $\Delta L_{min}$) and values of ${\cal C_M}$. Each setup is adopted at the merger time, while the inspiral phase is common to all of them and is run under the {\tt LR} configuration. The domain of the finest AMR grid for the {\tt MR} and {\tt HR} cases changes with time, so that the values indicated here are only approximated. We have also included a LES with MR and no magnetic fields ({\tt MR B0}) for comparison purposes.}
	\label{tab:models}
\end{table}

\begin{figure*}
	\centering
	\includegraphics[height=0.9\textwidth]{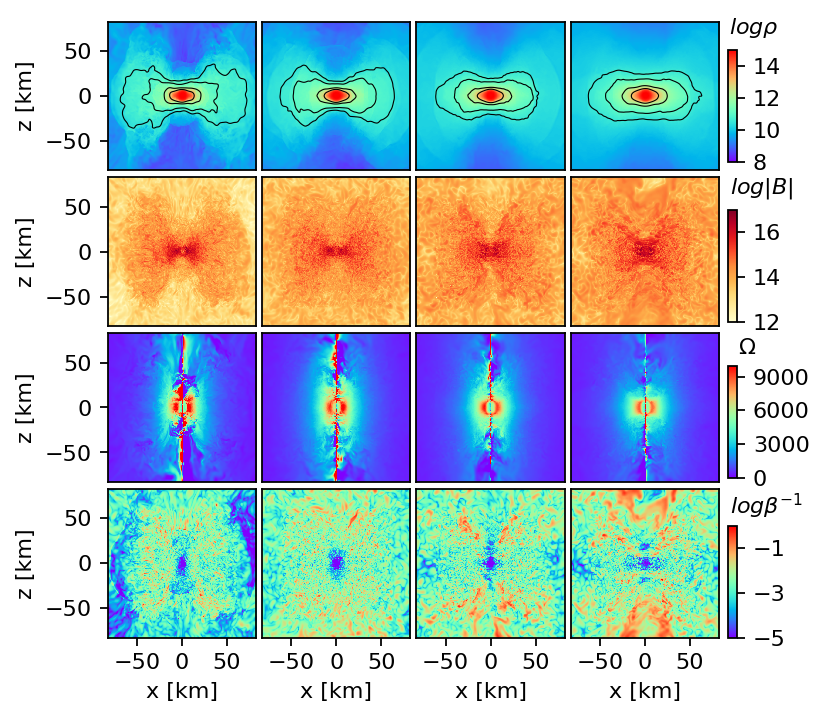}		
	\caption{{\em Evolution of relevant quantities of the remnant in a meridional plane.} The rows show, from top to bottom, the rest-mass density in $\mathrm{g}/\mathrm{cm}^{3}$, magnetic field intensity in Gauss, angular velocity of the fluid $\Omega$ in $\mathrm{rad}/\mathrm{s}$, and inverse of $\beta$ factor,
	at $t=(5,10,30,50) \,\mathrm{ms}$ after the merger, for the medium resolution LES simulation.  In the rest-mass density plots (first row), black solid lines represent constant density surfaces at $\rho=(5\times 10^{10},10^{11},10^{12},10^{13}) \mathrm{g}/\mathrm{cm}^{3}$. 
	}
	\label{fig:fields_meridional}	
\end{figure*}

Here we present a set of simulations of the equal-mass BNS initial data described before, with and without SGS terms for three different resolutions, as specified in Table~\ref{tab:models}.

\subsection{Qualitative dynamics}

The stars perform roughly $5$ orbits before merging. By inspecting the density profiles in the equatorial plane, we observe that after the first contact, the merging cores of the two stars perform radial bounces with a period of roughly 1 ms, embedded within a common envelope.
The cores perform roughly 10 damped oscillations before settling down into a slow-varying, approximately axisymmetric fluid configuration. Based on the density and angular velocity profiles, one can roughly separate the remnant into a high-density {\em bulk}, which extends for $\rho \geq 10^{13} \mathrm{g}/\mathrm{cm}^{3}$, and an {\it outer envelope} (a.k.a. disk) with densities $10^{13} \mathrm{g}/\mathrm{cm}^{3} \geq \rho \geq 5 \times 10^{10} \mathrm{g}/\mathrm{cm}^{3}$, which is initially torus-shaped and orbits close to Keplerian velocity. 
These regions quickly evolve with time, as shown in Figure~\ref{fig:density3D} for some representative snapshots during the first $10$~ms of the evolution. The bulk soon settles into a roughly ellipsoidal shape, with strong differential rotation, and an approximate equatorial radius of $15$~km. Similarly, the outer envelope region is very irregular and dynamical after the merger, but becomes more axi-symmetric and ellipsoidal-shaped at later times, covering a region that extends approximately to a radii of $50$~km in the equatorial plane. 

A general overview of the dynamics can be obtained from the various fields displayed in Figure~\ref{fig:fields_meridional}, showing the meridional $y=0$ plane. The first row shows the mass density distribution: soon after the merger, the remnant is surrounded by a torus-shaped distribution of marginally bound matter. At later times, the fluid in the torus expands in the polar directions, polluting the region near the $z$-axis. 
The magnetic field (second row) is amplified quickly in the few ms after the merger, and then it is spread by the fluid flow. The angular velocity of the fluid (third row) shows strong differential rotation near the transition between the bulk and the envelope, along both the cylindrical radial ($R$) and vertical ($z$) directions. Neither the shape nor the values of this quantity change significantly during the timescale of our simulations.  Finally, the evolution of $\beta^{-1}$ (fourth row) shows how the magnetic pressure gradually acquires importance compared to the fluid one. The pressure is always dominated by the fluid, except at late times near the rotational axis where they are comparable, suggesting the formation of a magnetically-dominated region.

\begin{figure*}
	\centering
	\includegraphics[width=1.0\textwidth]{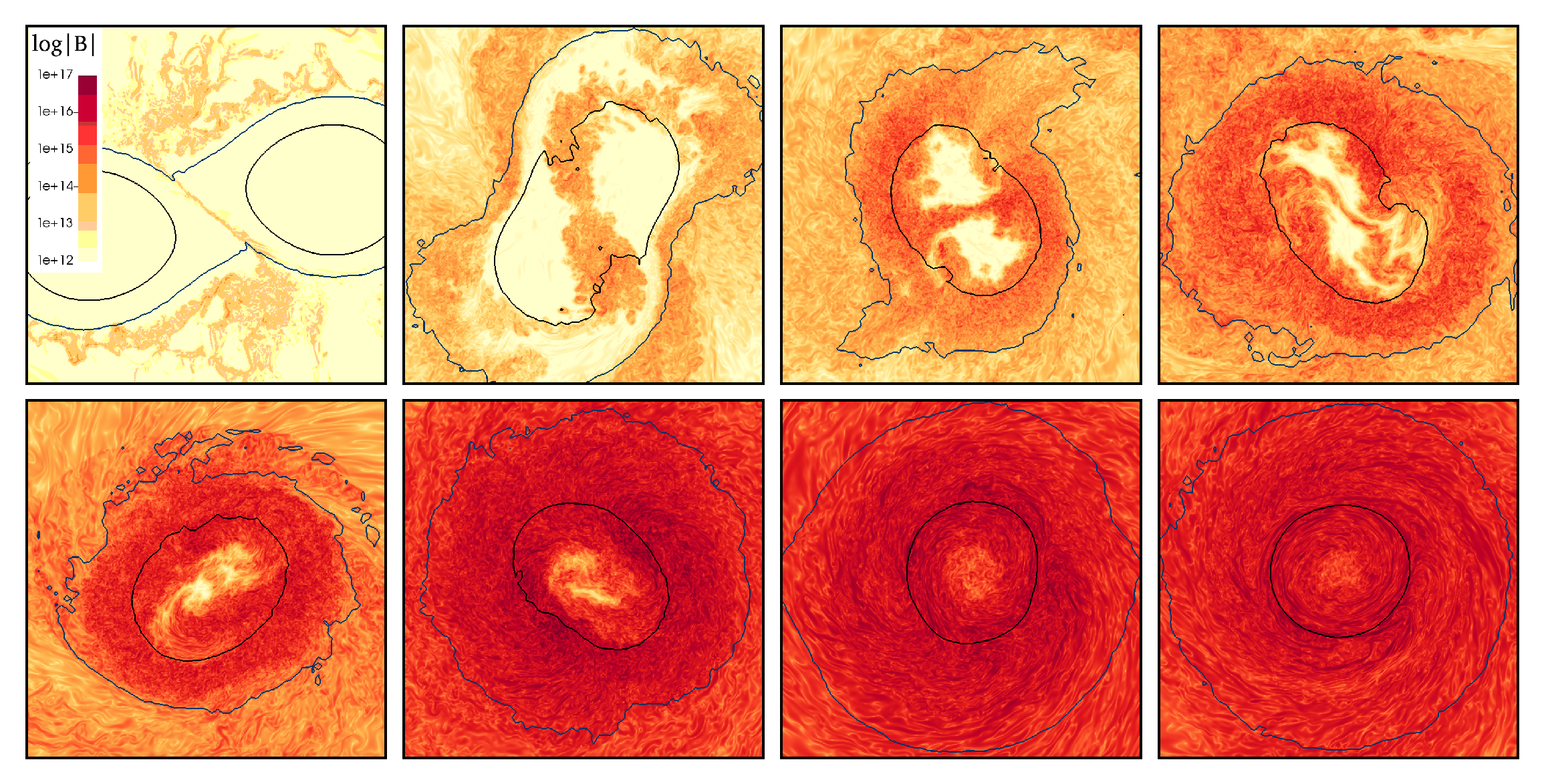}
	\caption{ {\em Evolution of the magnetic field intensity in the orbital plane}. Snapshots at $t=(0.5,1.5,2,2.5,3.5,5,10,15)$ ms after the merger, displaying the magnetic field strength in Gauss, together with constant density contours at $\rho=(10^{13},5\times 10^{14}) \mathrm{g}/\mathrm{cm}^{3}$  showing the transition region between the bulk and envelope of the remnant, as well as the location of very high dense regions. The amplification is mostly driven by the KHI, possibly with contributions from the RTI in the initial stages in the envelope (see top left plot). They induce a turbulent small-scale dynamo, generated mostly near the shear layer, that is quickly advected by the fluid motions and diffuses throughout both the bulk and the envelope.}
	\label{fig:Bfield_equatorial}
\end{figure*}	

Let us now focus on how the magnetic field amplification works in the first $10$~ms after the merger, along the lines of our previous discussion in Section~\ref{sec:picture}. The main features of the magnetic field evolution can be observed in Figure~\ref{fig:Bfield_equatorial}, where the field intensity and  iso-density contours at $\rho=(10^{13},5\times 10^{14}) \mathrm{g}/\mathrm{cm}^{3}$, identifying the transition between dense core, bulk and envelope, are displayed in the orbital plane for our highest-resolution run. A thin, rotating shear layer is produced at the time of the merger between the colliding cores that move in opposite directions. This is the perfect premise to develop the KHI, manifested by the typical curly structures at small scales. The fastest-growing modes are expected to be the ones with a wavelength similar to the layer thickness, which   might be much thinner (possibly, sub-meter scale)
than any currently achievable resolution. Our simulation is accurate enough  to capture the formation of very small eddies (measuring a few numerical points, see the finest details in color-scale the Figure~\ref{fig:Bfield_equatorial}).
In agreement with the general expectation, the non-linear interaction of these growing modes, with different scales, produces a turbulent dynamics with a direct energy cascade, where the kinetic energy is transferred from large to small scales. These small eddies twist and stretch the magnetic field of the remnant, increasing quickly the magnetic energy at the expenses of the kinetic one. As we will show later, this energy transfer is more efficient at small scales and acts as a powerful turbulent dynamo. There is also a significant growth of the magnetic field occurring at the surface of the stars during the merger, possibly due to a small-scale turbulence excited by the RTI or the KHI (i.e., see the top left panel in Figure~\ref{fig:Bfield_equatorial}).

After few miliseconds, the maximum intensity of the magnetic field lies not in the plane between the cores, but in the region separating bulk and outer envelope. The fluid mixing induced by the complex fluid flow during and after merger
is able to spread very efficiently the turbulent state of fluid and magnetic fields to most of the remnant in a short timescale. 
After reaching a peak, about $5$~ms after the merger, the magnetic field intensity is advected through the remnant, but it does not show any significant amplification during the first $20$~ms after the merger.
From that time until the end of our simulation, the dynamics develop larger and more coherent magnetic field structures whose intensity grow linearly with time, possibly due to the winding mechanism.

\subsection{Energetics evolution}

\begin{figure}
	\centering
	\includegraphics[width=0.5\textwidth]{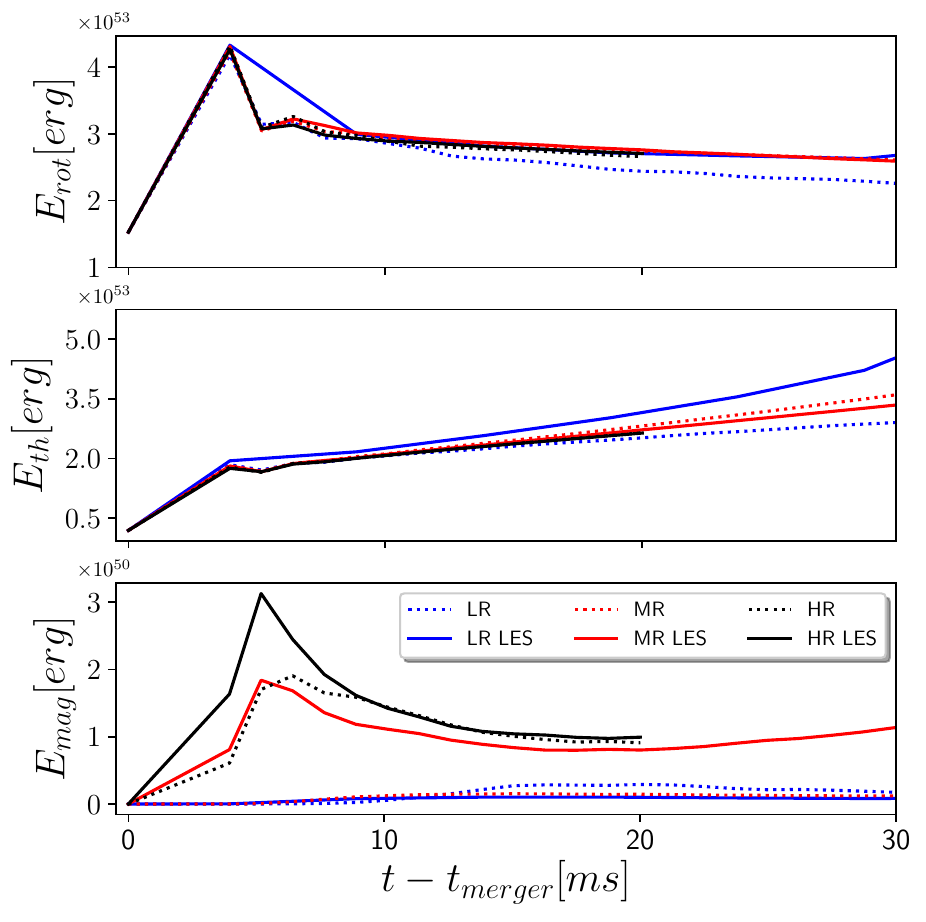}
	\caption{ {\em Evolution of the energies}. The panels show the rotational, thermal, and magnetic energies, integrated over the full domain. Colors identify the different resolutions, for LES (solid lines) and standard simulations (dashed). Note that these quantities are computed in a post-processing step from saved 3D data, which is available only every few ms for some portions of the simulations. This leads to an uneven time resolution of the curves, especially for the cases with less output like LR LES.
	}
	\label{fig:energies}
\end{figure}

In order to provide a more quantitative global analysis, we now discuss the evolution of the integrated energies and averaged fields in the remnant. The total rotational kinetic, thermal, and magnetic energies are displayed in Figure~\ref{fig:energies} for all our magnetized simulations. 
 
We observe that, around ${\sim}10$~ms after the merger, the rotational kinetic energy stabilizes at $E_\mathrm{rot} \sim 3 \times 10^{53}$~ergs. At this time, the remnant has settled down into a slowly-varying fluid configuration approaching axial symmetry. The thermal energy is of a similar order and slowly increases in time. This is likely due to shock heating, first in the merging stars, and later in the envelope, which is perturbed by the oscillating remnant. 
For comparison, the total baryonic mass (which is conserved up to $0.7\%$ during the time span by our simulations) corresponds to an energy of about $E_\mathrm{bar} \sim 5 \times 10^{54}$~ergs.

A comparison between different resolutions shows that in the low-resolution case (blue curves) a significant part of the heating comes from the intrinsic dissipation of the numerical scheme (as on can infer from the faster increase). For the other cases (LES and standard simulations) the differences are minor, indicating that (i) they are in the convergence regime and (ii) the LES approach with the SGS gradient model does not introduce spurious heating in the fluid evolution. On the other hand, the magnetic energy evolution shows significant differences between the two groups of simulations. Both high-resolution simulations (LES and standard) and the medium-resolution LES all present a sharp increase in the first few milliseconds after the merger. After $5$~ms, the magnetic energy  reaches a peak $E_\mathrm{mag} \sim 3\times 10^{50}$~ergs, three orders of magnitude below the thermal and rotational energy. During the next $15$~ms, the magnetic field energy slightly decreases and saturates roughly at $10^{50}$~ergs. Then it increases again, following a quadratic function, and reaching $E_\mathrm{mag} \sim 2\times 10^{50}$~ergs towards the end of our longest simulation, at $50$~ms after the merger. In contrast, the low-resolution simulations (LES and standard) and the medium-resolution standard simulation are incapable of capturing the instability that induces this magnetic field amplification.
At this point we would like to recall that the SGS terms are not being applied in the outer envelope, so our results demonstrate the benefits only in the bulk region.

\subsection{Convergent magnetic field amplification: the effects of KHI, winding and MRI}

\begin{figure}
	\centering
	\includegraphics[width=0.48\textwidth]{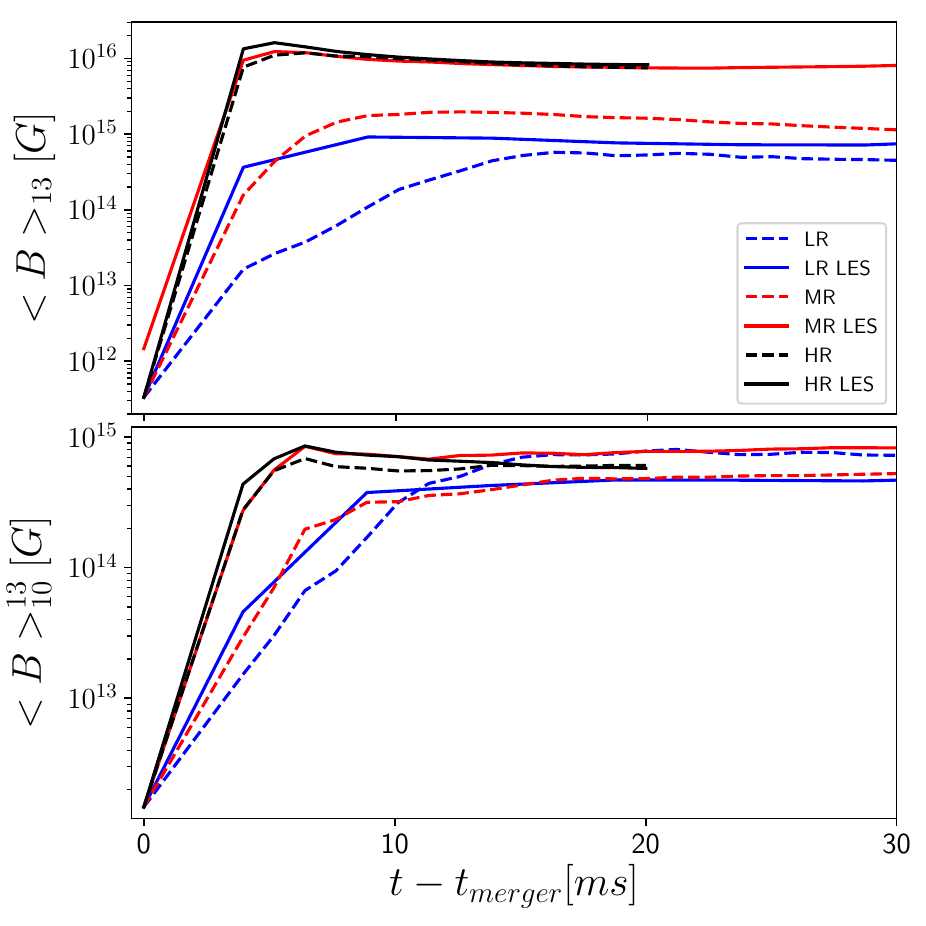}	
	\caption{ {\em Evolution of the average magnetic field.} Volume-average of the magnetic field intensity either in the bulk (top) or in the envelope (bottom) of the remnant. These averages are calculated for all the resolutions (identified by colors), both for the LES (solid lines) and for the standard simulations (dashes). Clearly, the averages in the bulk calculated with LES converge to a well-defined value.}
	\label{fig:averageB_all}
\end{figure}

To get a more fine-grained picture of the magnetic amplification, we now study the averages in the bulk and the outer envelope separately. 
In Figure~\ref{fig:averageB_all}, we show the magnetic field intensity for all the simulations, averaged either in the bulk or in the envelope. In the bulk, the averaged intensities differ by up to two orders of magnitude between the lowest and the highest resolutions. For all of the cases, magnetic fields are amplified by at least three orders of magnitude in a short timescale lasting about $5$~ms. This amplification phase is followed by a saturation phase where the intensity is nearly constant for tens of milliseconds. The standard simulations display the lack of convergence commonly found already in the literature; as the resolution is increased, so does the saturation level reached by the averaged magnetic field. 

The LES with the gradient SGS model behave in a completely different manner. For the first time, to our knowledge, we found convergence in the evolution of the averaged magnetic field: the medium resolution LES already grows and saturates at the same quantitative level, approximately of $10^{16}$~G, as the high-resolution LES. Interestingly, this value is almost identical to the one achieved by the standard high resolution simulation,  suggesting that the magnetic field already converged also in this case. Unfortunately, the computational cost of standard simulation with even higher resolution prevented us of performing additional tests to confirm this result.
By using LES with the gradient model, the simulations are apparently able to capture most of the effects of the small-scale, unresolved dynamics during the turbulent stage. Note that even with the LES, high resolution and high-order numerical schemes are required  in order to reach a faithful saturation level.

In the envelope, all the simulations reach similar values $\sim 5\times 10^{14}-10^{15}$~G at saturation. However, we stress that in most of the envelope the resolution is fixed and the SGS terms are disabled for all these cases.
We recall that the focus of our study is to test  the convergence of the magnetic field amplification within the bulk. The differences in the envelope are mainly caused by the differences within the bulk. We find that the latter is not causing large differences in the envelope. 
The aforementioned limitations (i.e., fixed resolution and no SGS terms) prevent us from drawing further conclusions about convergence of the magnetic field within the envelope, which may be subject of future studies.

Let us note that the divergence of the magnetic field, normalized by the magnetic field strength, remained relatively small ${\cal O} (0.005)$ during all the simulation for both the standard simulation and the LES with the gradient SGS term. The fact that these normalized values  are almost the same suggests that the violations of the solenoidal constraint does not interact in an unexpected way with the gradient SGS term to induce a fictitious amplification of the magnetic field.
%

\begin{figure}
	\centering
	\includegraphics[width=0.48\textwidth]{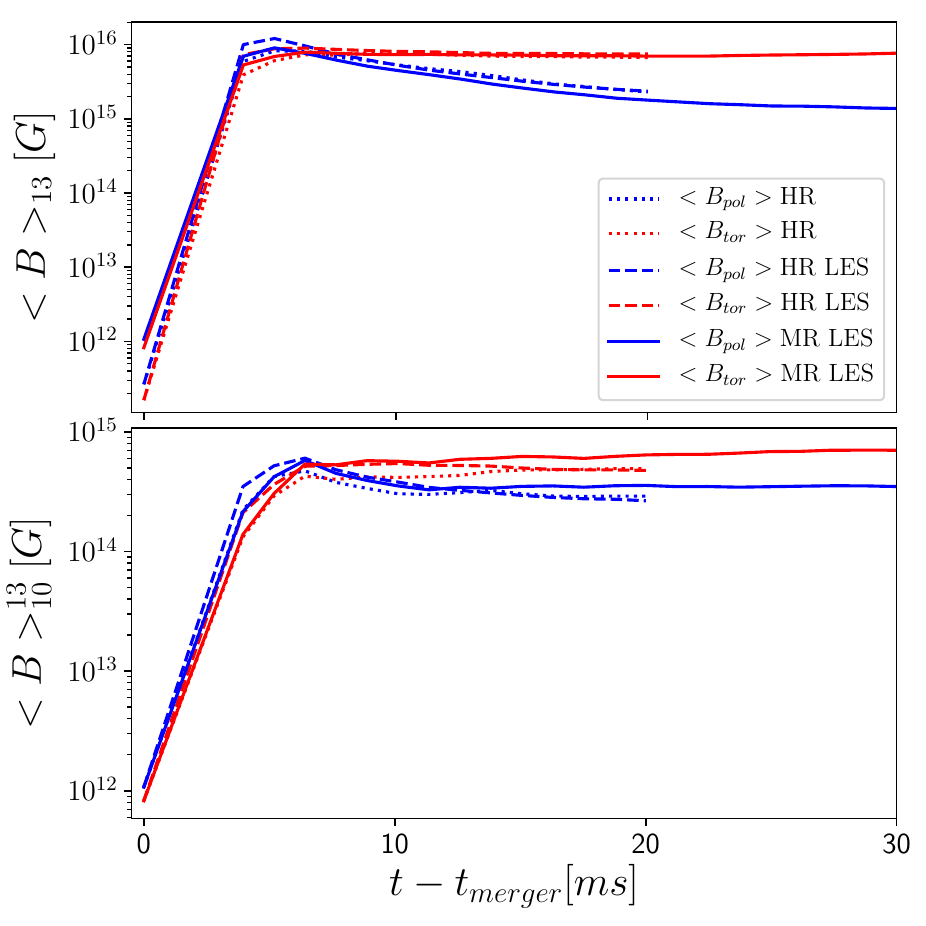}	
	\caption{ {\em Evolution of the average poloidal and toroidal magnetic field components}. Contributions of the poloidal (blue) and toroidal (red) components to the magnetic field intensity, volume-averaged in the bulk (top) or in the envelope (bottom), for the medium resolution LES and the high-resolution simulations (linestyles).}
	\label{fig:averageB_components}
\end{figure}

In Figure~\ref{fig:averageB_components} we illustrate the poloidal and the toroidal contributions to the average magnetic field, for the high-resolution simulations and the medium resolution LES. First, note that these three simulations approach the same solution, which shows that the numerical dissipation does not dominate the dynamics, and demonstrates that solution obtained with medium-resolution LES is already a good approximation to the high-resolution one. During the kinematic phase, i.e. while the magnetic field's feedback on the fluid is dynamically irrelevant, both components grow exponentially at a similar rate. This is a strong indication that such amplification is due to small-scale turbulent dynamo induced by the KHI (and possibly the RTI), which is naturally isotropic, providing amplified magnetic fields in random directions. This stage ends when the magnetic fields are strong enough to back-react significantly on the fluid, leading to a quasi-stationary state (in average) at saturation. During this last stage, the toroidal magnetic field, averaged in the bulk, remains roughly constant at values $\sim 10^{16}$~G until $20$ ms and then it starts to grow linearly. The poloidal component, however, decreases almost by an order of magnitude until 30 ms. This qualitative behavior is also reproduced in the envelope, although the difference between both components is less significant (i.e., just a factor of a few) in that region.

\begin{figure}
	\centering
	\includegraphics[width=0.45\textwidth]{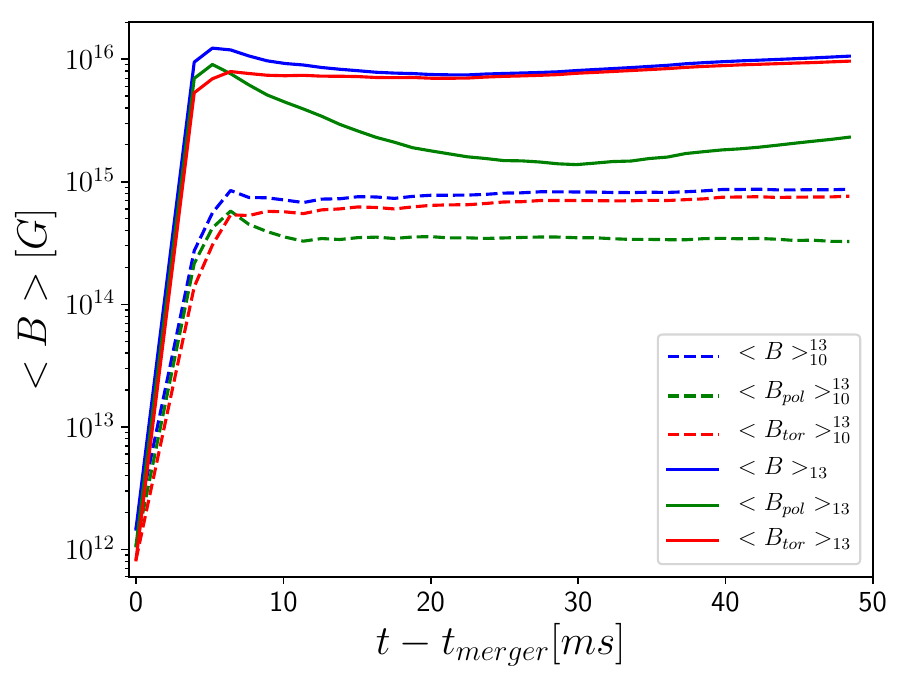}
	\caption{ {\em Evolution of the average magnetic field (and components) in long timescales}. Averaged magnetic field intensity and its components, considering the bulk (solid) or the envelope (dashes), for the medium resolution LES. After the amplification growth both component are comparable,
	the expected outcome for the saturation state in isotropic turbulence. The poloidal component decreases until $t\sim30$ ms while the toroidal one remains constant. At later times, both components grow linearly (i.e., the toroidal component growth two times faster than the poloidal one), probably due to the winding mechanism.
	}
	\label{fig:averaged_MRLES}
\end{figure}

After validating the medium resolution accuracy via comparisons between resolutions, we have continued the medium-resolution LES up to timescales much longer than feasible with the high resolution. The total averaged magnetic field and its components are displayed in Figure~\ref{fig:averaged_MRLES} up to $50$~ms, showing the main features described before. After $30$~ms the poloidal component also grows within the bulk. Although there can be a partial contribution from the MRI, the linear growth in both magnetic field components strongly suggest that it can be mostly attributed to the winding mechanism, acting more efficiently in the  spherical-shaped region of strong differential rotation.
Interestingly, the toroidal component grows at a rate two times faster than the poloidal one.

\begin{figure}
	\centering
	\includegraphics[width=0.45\textwidth]{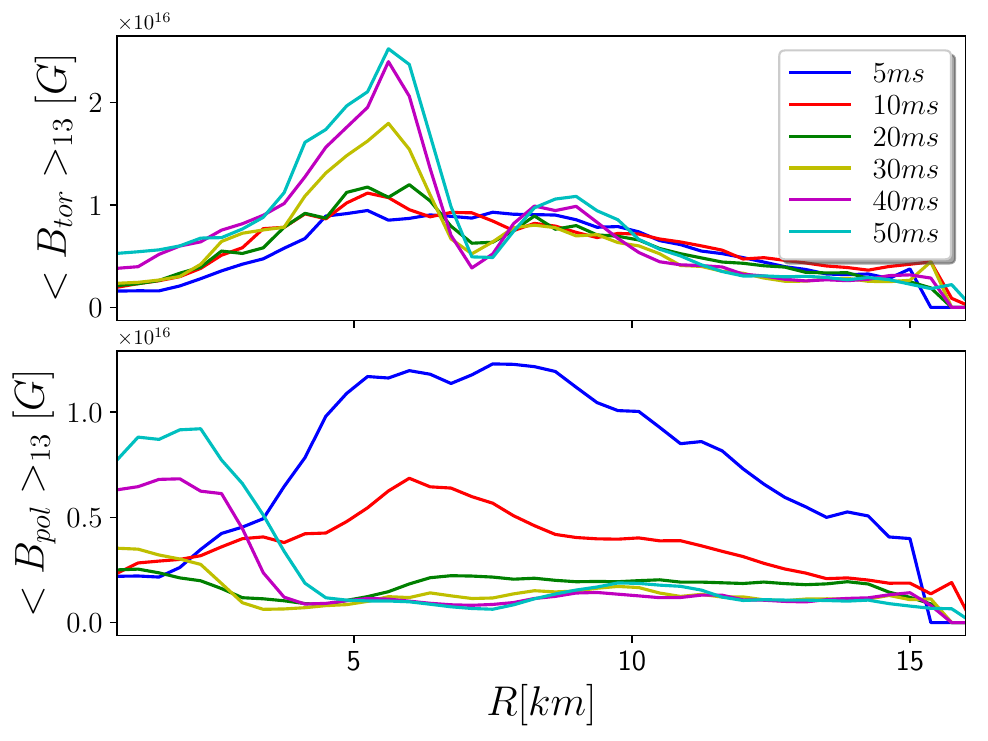}
	\includegraphics[width=0.45\textwidth]{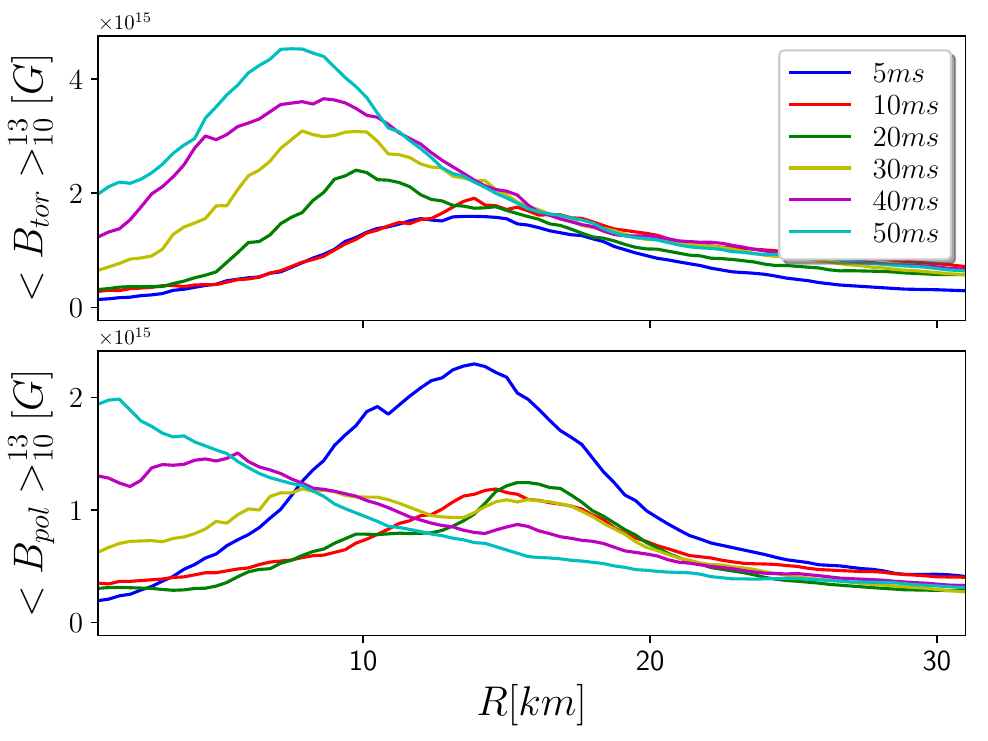}	
	\caption{ {\em Evolution of the (cylindrical) radial distribution of the magnetic field components}. The toroidal and poloidal components of the magnetic field, averaged at different cylindrical surfaces $R$ either in the bulk (top) or in the envelope (bottom), for the medium-resolution LES. At later times both components grow in regions $R \lesssim 6$ km in the bulk and $R \lesssim 15$ km in the envelope.}
	\label{fig:B_radial}	
\end{figure}

As a complementary perspective, Figure~\ref{fig:B_radial} shows the (cylindrical) radial profiles of the toroidal and poloidal components of the magnetic field, averaged on cylinder surfaces (i.e., in the azimuthal and vertical directions), at different times. Again, we perform these averages separately in the bulk and in the envelope. As it was observed in the volume average, both components grow in a broad spatial region to values ${\cal O}(10^{16})$~G in the bulk and ${\cal O}(10^{15})$~G in the envelope during the first $5$~ms after the merger, likely due to the KHI amplification and the fluid mixing of the turbulent state. Then, in the next $15$~ms, the poloidal component decreases and get flattened, while that the toroidal component remains almost unaltered. In the following $30$~ms there is a growth of the toroidal component in the bulk around $R \sim 6$~km and in the envelope around $R \sim 10$~km. This roughly linear growth is followed, with a time delay of approximately $10$~ms, by a similar one of the poloidal component in the interior of the peak developed by the toroidal component.

Notice that the magnetic field amplification occurs in regions where MRI cannot be excited because $d\Omega/dR > 0$. These two features strongly suggest that this growth is indeed produced by the winding mechanism, acting near the transition between the bulk and the envelope, where the differential rotation displays a maximum. By combining the profiles from both components is clear that the magnetic field is evolving towards a helical  configuration similar to those present in magnetically dominated jets.

\subsection{Spectral distribution}

\begin{figure*}
	\centering
	\includegraphics[width=0.9\textwidth]{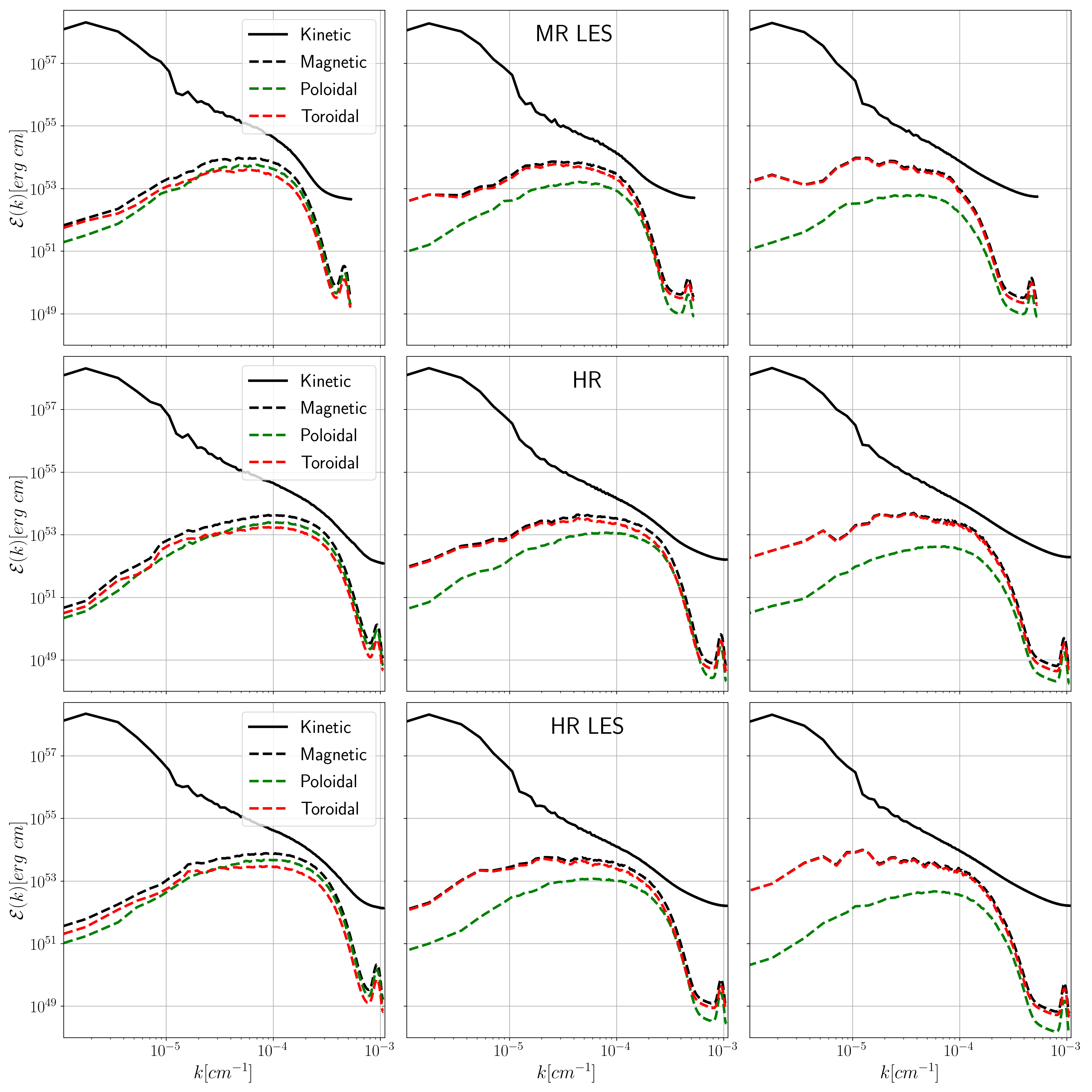}
	\caption{ {\em Evolution of the energy spectra}. Spectra of the kinetic (solid) and magnetic energy (dashed) at $t=(5,10,20)$ ms, for the medium resolution LES and the high-resolution simulations. The kinetic and magnetic energies almost reach equipartition at high wavenumbers (i.e., before the decay induced by numerical dissipation, which is much more prominent in the magnetic spectra). Note that the  high-resolution simulations agrees very well with the medium resolution LES, maybe except for low wavenumbers where the statistical error of the spectra is larger and, in any case, whose modes represent a very small fraction of the energy} .
	\label{fig:energy_spectra}		
\end{figure*}

A more comprehensive information of the energy distribution over the spatial scales can be obtained from the power spectrum. The kinetic and magnetic spectra (eqs. (\ref{eq:spectra_k})-(\ref{eq:spectra_b})) for the high resolution simulations and the medium-resolution LES are displayed in Figure~\ref{fig:energy_spectra} at times $t=(5,10,20)\,\mathrm{ms}$. Here we can observe some of the features described before, like the isotropic turbulent state driven by the KHI that produces a similar spectra of the poloidal and toroidal components at $5\,\mathrm{ms}$. The magnetic energy approaches equipartition with the kinetic one at high wavenumbers $k$~\footnote{Notice that at the highest wavenumbers, spectra are damped by the numerical dissipation of the HRSC numerical scheme. In that region there is also a pile-up of magnetic energy due to the non-periodic boundary conditions of the domain where the spectra is calculated.
}.

The kinetic spectra is well represented by the Kolmogorov power law $k^{-5/3}$, the typical direct cascade of isotropic kinetic turbulence. The magnetic field at low wavenumbers follow the Kazantsev power law $k^{3/2}$, as expected for non-saturated small-scale turbulent MHD dynamo. At later time, the overall poloidal spectra decreases slightly, as we already noticed in the averages described previously. The spectra of the toroidal component increases, with a clear transfer from large wavenumbers to small ones (i.e., from small to large scales), indicating the presence of a inverse energy cascade. It is very important to stress that, already in these short timescales, the peak of the magnetic field spectra moves to lower wavenumbers. The fact that the toroidal component is the most affected indicates again that the winding is probably the responsible for this energy transfer.

\begin{figure*}
	\centering
	\includegraphics[width=0.95\textwidth]{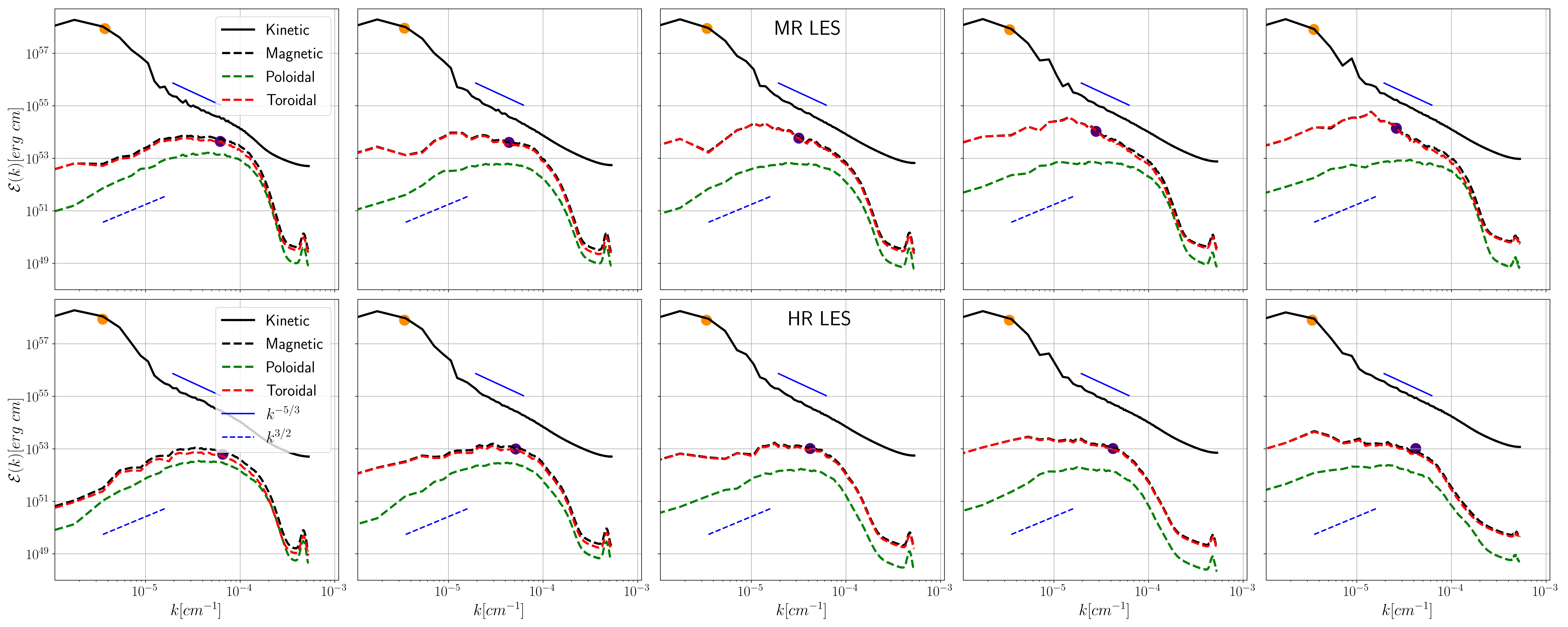}
	\caption{ {\em Evolution of the energy spectra at late times}. 
	Kinetic and magnetic spectra for the medium resolution cases at $t=(10,20,30,40,50)\,\mathrm{ms}$. The standard medium resolution simulation is only qualitatively similar to the medium resolution LES. As expected in turbulence with low magnetization, at $t=10$~ms the kinetic energy spectra shows the standard Kolmogorov power law $k^{-5/3}$ (short solid blue line)  in the inertial range, while the magnetic energy follows the Kazantsev power law $k^{3/2}$ (short dashed blue line) at low wavenumbers. At late times, the magnetic energy spectra follows the Kolmogorov slope at intermediate and high wavenumbers. The weighted-average wavenumbers ${\bar k}$ are represented by orange and black dots, corresponding to the typical coherent scales of fluid and magnetic structures.
	}
	\label{fig:energy_spectra_long}			
\end{figure*}

We have followed the evolution of the kinetic and magnetic spectra at later times for the medium resolution simulations with and without SGS terms, as shown in Figure~\ref{fig:energy_spectra_long}. 
Even on longer timescales up to 50~ms, the kinetic energy spectra remains mostly unchanged for both simulations. However, there are significant differences in the magnetic spectra. In the medium-resolution LES, the peak of the magnetic energy spectra moves from high to intermediate wavenumbers (i.e., from small to intermediate spatial scales). The magnetic energy distribution, as time progresses, maintains roughly the Kazantsev power law on large scales and the Kolmogorov power law at small ones. 

We also calculate, for each case, the spectra-weighted average wavenumber ${\bar k}$ and its associated lengthscale $\delta R = 2\pi/{\bar k}$, which roughly represents the distance at which the field is not randomly oriented but tends to show a defined direction. For the medium resolution LES, $\delta R$ of the magnetic energy moves from $0.7$~km to $2$~km during the $50$~ms of the simulation. On the other hand, $\delta R$ of the kinetic energy varies only from $12$~km to $17$~km. The standard medium resolution simulation share roughly the same qualitative behavior, but with a magnetic energy spectra much more flattened and many orders of magnitude smaller.

\subsection{Differential rotation and relevant profiles near the rotation axis}

\begin{figure}
	\centering
	\includegraphics[width=0.45\textwidth]{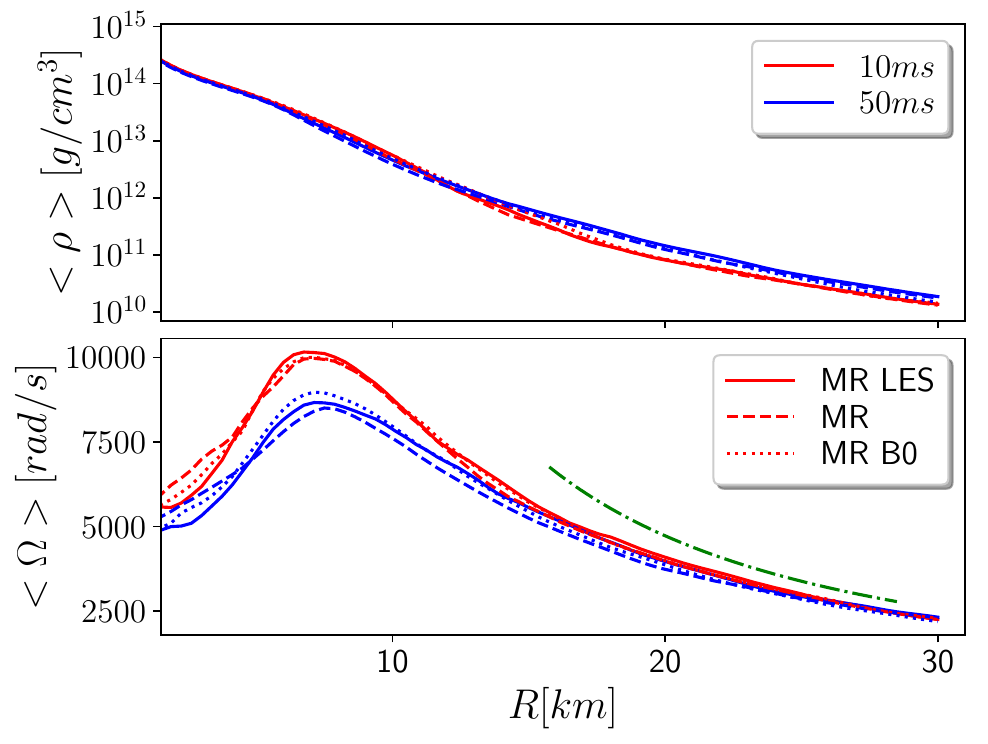}
	\caption{ {\em Evolution of the radial distribution of density and angular velocity}. Profile of the density and $\Omega$ in the orbital plane, averaged over the azimuthal direction. Two different times (colors), $t=(10,50)$ ms are represented for the medium resolution LES (solid) and the standard simulation (dashed). We also include the non-magnetized case {\tt MR B0} (dotted), and show the falloff behavior (green) $\Omega_K \sim R^{-3/2}$ of Keplerian disks for comparison.}
	\label{fig:density_omega_radial}
\end{figure}

Here we turn our attention to a more detailed analysis of the differential rotation profile present in the remnant, comparing the simulations with the same medium resolution, standard and LES. In order to disentangle the effects from the magnetic field, we also performed a standard simulation with no magnetic field. The radial profiles of density $\rho(R)$ and angular velocity $\Omega(R)$, for two different times $t=(10,50)$~ms after merger, are displayed in Figure~\ref{fig:density_omega_radial}. These profiles are extracted by computing the azimuthal averages within the orbital plane (i.e., along a circle in the $z=0$ plane). 
The first observation is that the density profile barely changes after the remnant has settled down into a roughly axisymmetric shape. The differences between the three different simulations are minor, indicating that magnetic field effects on the density profile are rather small. 
The angular velocity, at 10\,ms post-merger, shows in all three cases a radial profile with values $\approx\,6000$\,rad s$^{-1}$ near the orbital axis, increasing up to $\approx\,10000$\,rad s$^{-1}$ at $R \approx\,7.5$\,km. At larger distances it decreases, with falloff behavior approaching that of  Keplerian profiles (represented for comparison with a green dashed line in the plot). As time evolves, the profiles get more flattened, with the maximum $\Omega(R)$ decreasing to $\approx\,8500$\,rad s$^{-1}$ at 50\,ms post-merger. 
Comparing the three cases, we find that the amplified magnetic fields do not seem to have a significant impact on the removal of differential rotation.

\begin{figure}
	\centering
	\includegraphics[width=0.45\textwidth]{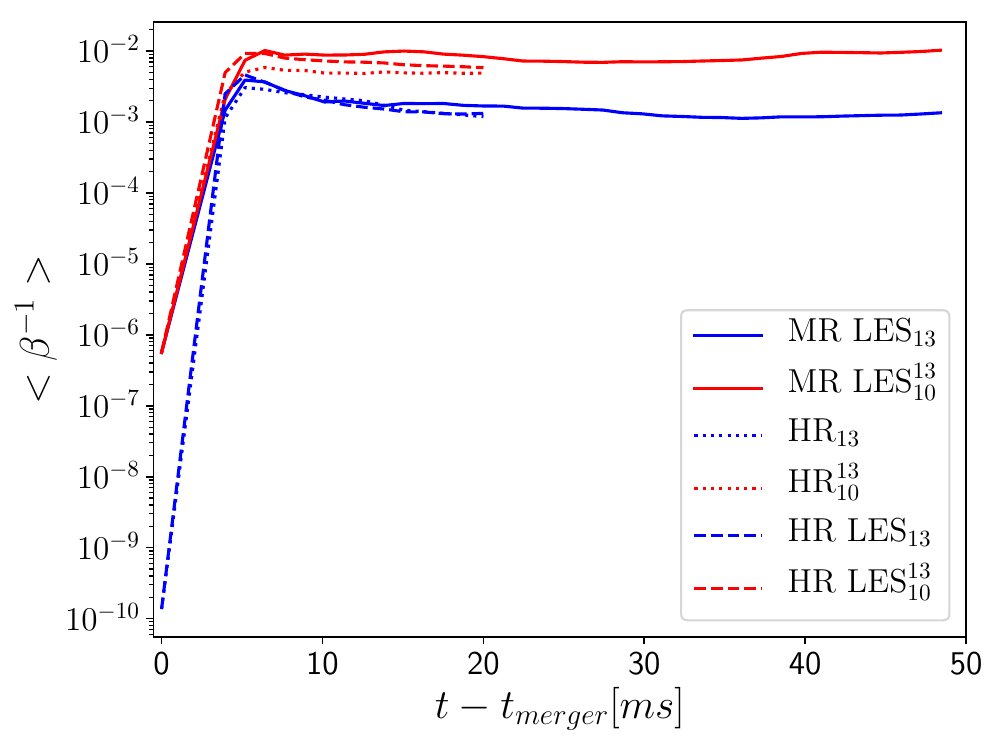}
	\caption{ {\em Evolution of the inverse of plasma beta}. Volume-averaged inverse of plasma beta, ${<}\beta^{-1}{>} = {<}B^2 / (2 P){>}$, within the bulk and within the envelope, for the medium resolution LES and the high-resolution simulations.}
	\label{fig:invbeta}
\end{figure}

In order to monitor the formation of magnetically dominated regions, we use the averages of $\beta^{-1}$, which estimate the relative importance of the magnetic pressure with respect to the fluid pressure. Figure~\ref{fig:invbeta} shows the volume averages in the bulk and in the envelope, for the high-resolution and the medium-resolution LES. On average, the magnetic pressure reaches about $1\%$ of the fluid pressure within the bulk, and 10 times less within the envelope. Again, the high-resolution simulations show a very similar behavior.

\begin{figure}
	\centering
	\includegraphics[width=0.45\textwidth]{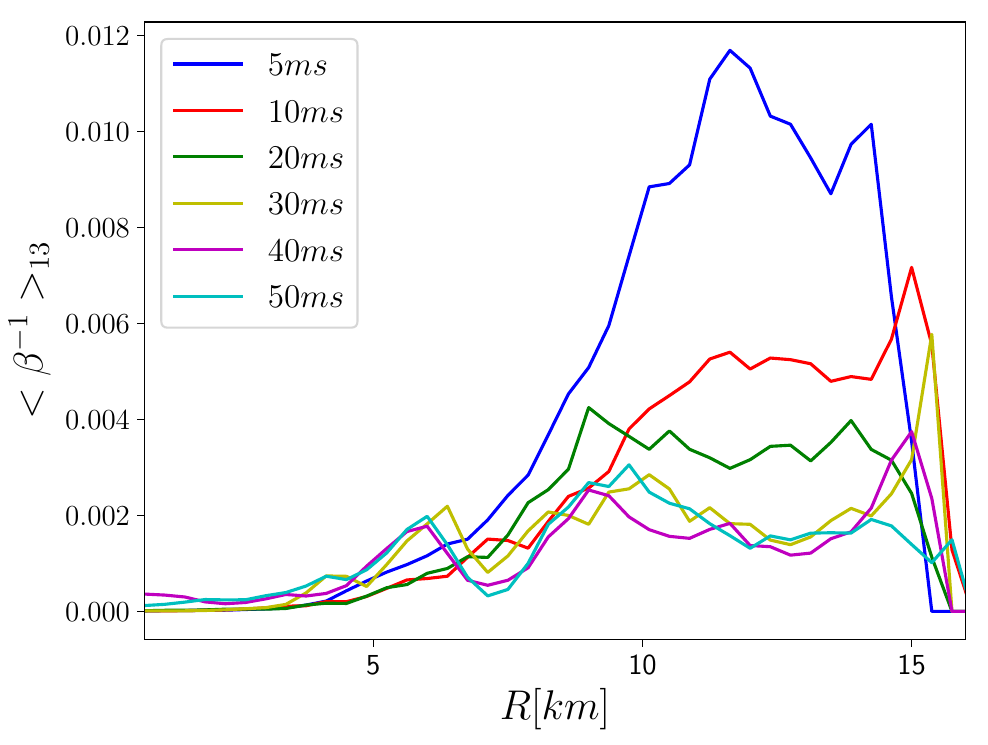}
	\includegraphics[width=0.45\textwidth]{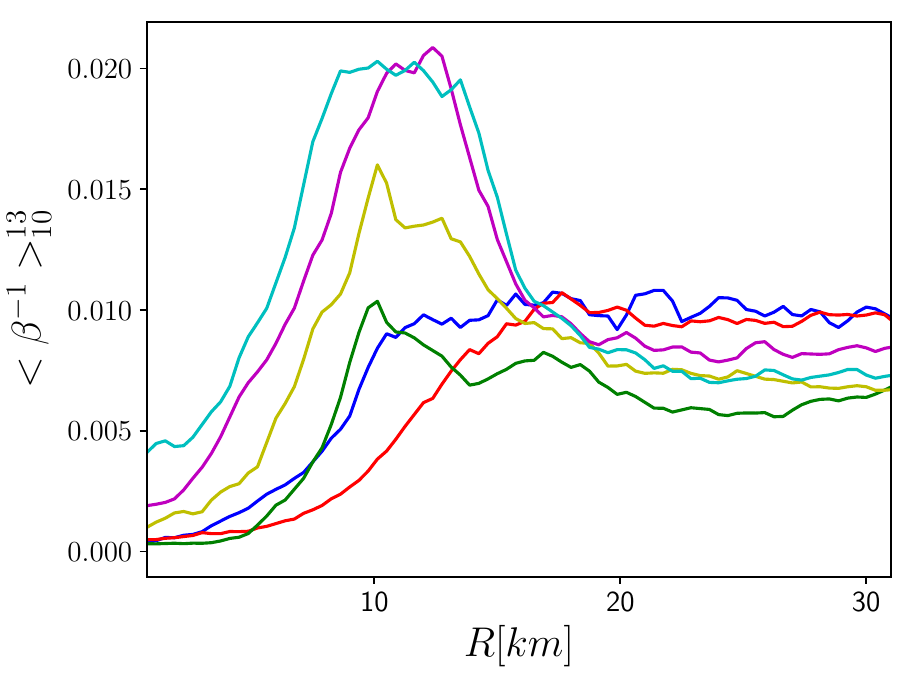}	
	\caption{  {\em Evolution of the radial distribution of $\beta^{-1}$}. The profile of the inverse of beta, averaged at different cylindrical surfaces $R$ either in the bulk (top) or in the envelope (bottom), for the medium-resolution LES. At later times the magnetic pressure becomes more relevant in the envelope for $R \leq 15$ km.
    }
	\label{fig:invbeta_radial}		
\end{figure}

We also calculate the averages of this quantity over cylindrical surfaces, as a function of time, to better understand the dynamical evolution. The field $\beta^{-1}$ averaged in the bulk, shown in the top panel of  Figure~\ref{fig:invbeta_radial}, reaches a peak at $t=5$~ms after the merger in the region $R \sim 11$~km. As time progresses, the profile  gets flattened, however, decreasing by almost an order of magnitude. This is probably related to the decrease of the poloidal magnetic field, due to rotation, once the saturation stage is achieved. Interestingly, the magnetic pressure becomes more relevant at later times in the envelope. At $t=5$~ms there is a broad distribution with $\beta^{-1} \sim 0.01$, almost flat for $R \sim 11$~km, but monotonically decreasing to zero at smaller radii. At $t=30$~ms after the merger, a bump appears in the profile around $R \sim 11$~km. Although the peak of this bump is only a factor 2 larger than the value in the flat region, it evolves towards smaller radii, increasing significantly the relevance of magnetic pressure  near the $z$-axis. This is another indication that magnetic field is growing  near the rotation axis of the remnant. 

\begin{figure}
	\centering
	\includegraphics[width=0.50\textwidth]{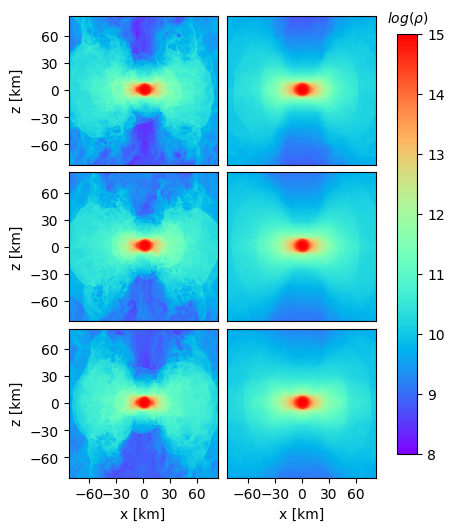}
	\caption{{\em Density in a meridional plane.} Density at $t=5$ ms (left panels) and $50$ ms (right) after the merger, for the medium resolution simulations: (top) without magnetic field, (middle)  standard simulation, and (bottom) LES including the SGS terms. Clearly, stronger magnetic fields increase the matter contamination of the funnel region near the orbital axis.}
	\label{fig:density_meridional}	
\end{figure}

\begin{figure}
	\centering
	\includegraphics[width=0.45\textwidth]{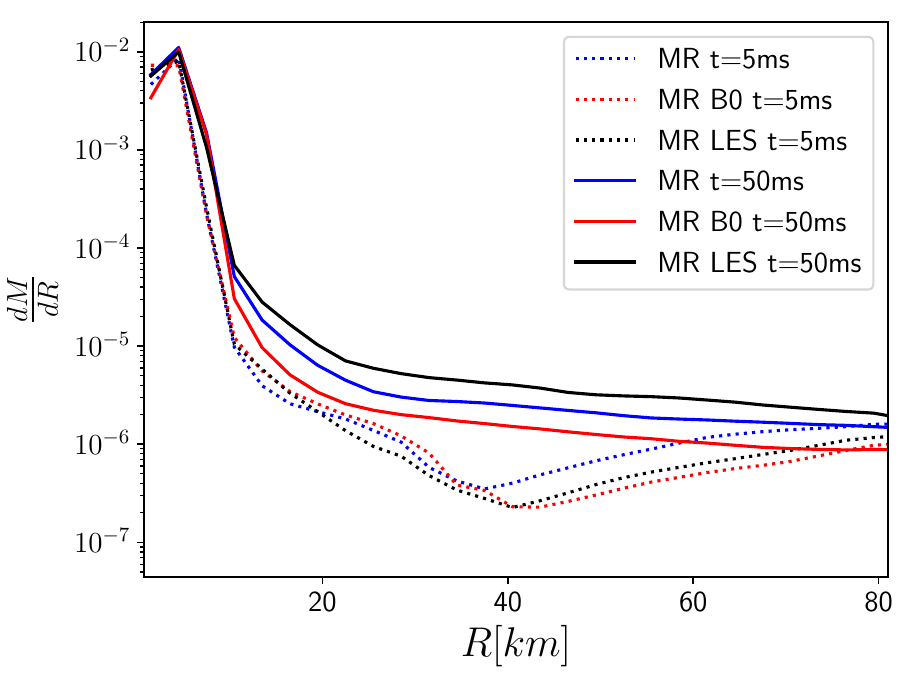}
	\caption{ {\em Distribution of density in a solid angle near the axis}. The mass contained in a solid angle of $15^{\circ}$ around the $z$-axis, for the medium-resolutions at $t=5$~ms and $t=50$~ms after the merger. This plot represents how the radial distribution of the mass increases with the magnetic field.
	}		
	\label{fig:histogram_15}	
\end{figure}

Near the rotation axis there are additional effects which might actually prevent the formation of magnetically dominated structures. Fig.~\ref{fig:density_meridional} displays the density in a meridional plane for the three medium resolution simulations: LES, standard and without magnetic field. There is a clear trend; the larger the magnetic field, the more dense is the region near the rotation axis. This effect can be quantified by computing the mass distribution in a solid angle of $15^{\circ}$ near the axis, as a function of the distance from the center, as it is displayed in Fig.~\ref{fig:histogram_15}. Despite starting from a similar distribution at $t=5$ ms after the merger, the mass distribution in the envelope increases faster for stronger magnetic fields, leading to a difference of approximately a factor three between the medium resolution LES and the non-magnetized simulation. This enhanced baryon contamination near the $z$-axis due to magnetic fields makes the possible emergence of a jet less likely~\cite{ciolfi2019,ciolfi2020collimated}.

\subsection{Gravitational waves and ejecta}

\begin{figure}
	\centering
	\includegraphics[width=0.45\textwidth]{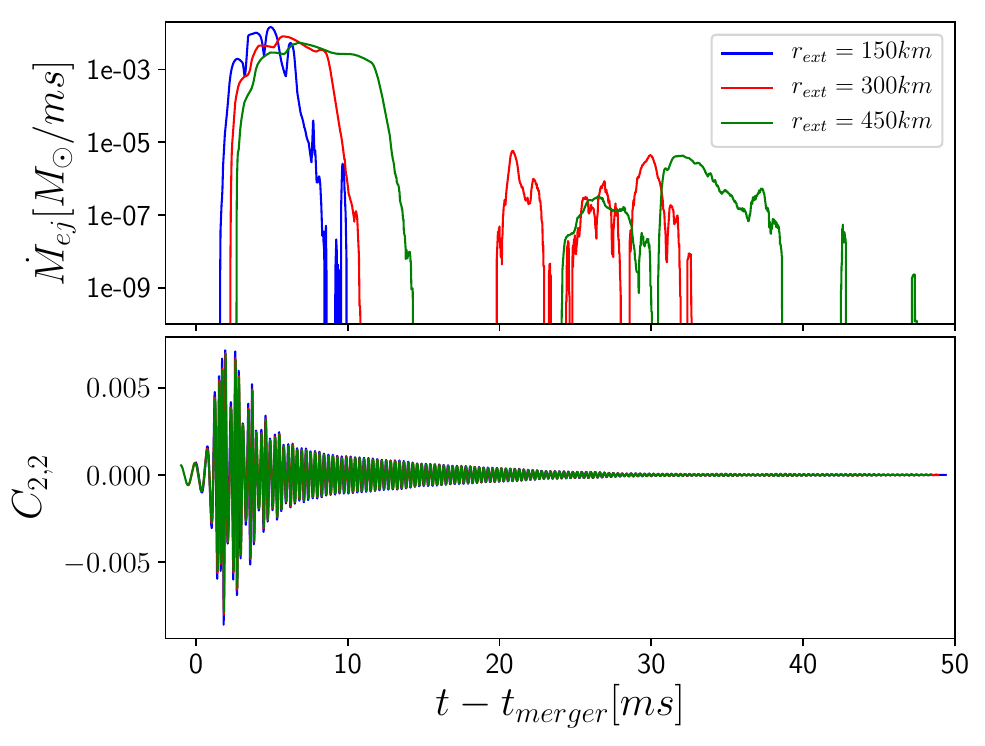}
	\caption{ {\em Unbound mass flux and gravitational waveform at different extraction surfaces}. Flux of unbound mass (top) and dominant mode (i.e.,$l=m=2$) of the Newman-Penrose $\Psi_4$ (bottom), as a function of time. Both quantities are computed at the three spherical surfaces located at $r_{ext}=(150,300,450)$~km.
	}
	\label{psi4_fluxmass}
\end{figure}

Finally, we discuss some aspects related to matter ejection and GW emission by the remnant.
Ejected matter can be split in two categories. The dynamical ejecta are launched during merger, as a result of tidal interactions and shocks formed in the neutron stars. Perturbations of the remnant can also liberate matter within tens of milliseconds after merger. The secular ejecta are emitted from the outer layers of the remnant on longer timescales by effects such as effective magnetic viscosity, magnetic pressure, neutrino radiation, or nuclear heating.
The flux of unbound mass observed at different distances is shown in the top panel of Figure~\ref{psi4_fluxmass}. By integrating this flux in time, we obtain an estimate for the total ejected mass $M_{ej}\approx 0.025 M_{\odot}$. 
As one can see, the ejecta in our simulations are dynamical, whereas
the secular ejecta contribution is negligible.
However, a further substantial mass outflow may be expected at later times. In Ref.~\cite{CiolfiKalinani20}, the authors studied the magnetically driven baryon-loaded wind launched in the post-merger for a similar BNS model (although with much lower resolution), finding a massive outflow emerging around the time of saturation of the magnetic energy growth ($\approx\!50$\,ms) and lasting up to 200\,ms.
We also note that the mass outflow may be partially hampered by the braking effect of the floor density in the atmosphere.
To assess the error caused by extraction at finite radii, we compare the total ejected mass for three different radii. The differences obtained are smaller than $3\%$ of the total value.

The dominant gravitational wave component is displayed in the bottom panel of Figure~\ref{psi4_fluxmass} for the medium resolution LES. 
To assess the error due to finite extraction radius, we show the signal calculated at three different extraction radii, demonstrating good agreement.
During the first $15$~ms, the gravitational wave signal exhibits a rich structure typical for merger remnants,
with strongly varying frequency and amplitude.
As the remnant settles down, the signal relaxes to a simple decaying oscillatory function. At the end of our simulation, at $t=50$~ms, the signal is almost zero. 

\begin{figure}
	\centering
	\includegraphics[width=0.45\textwidth]{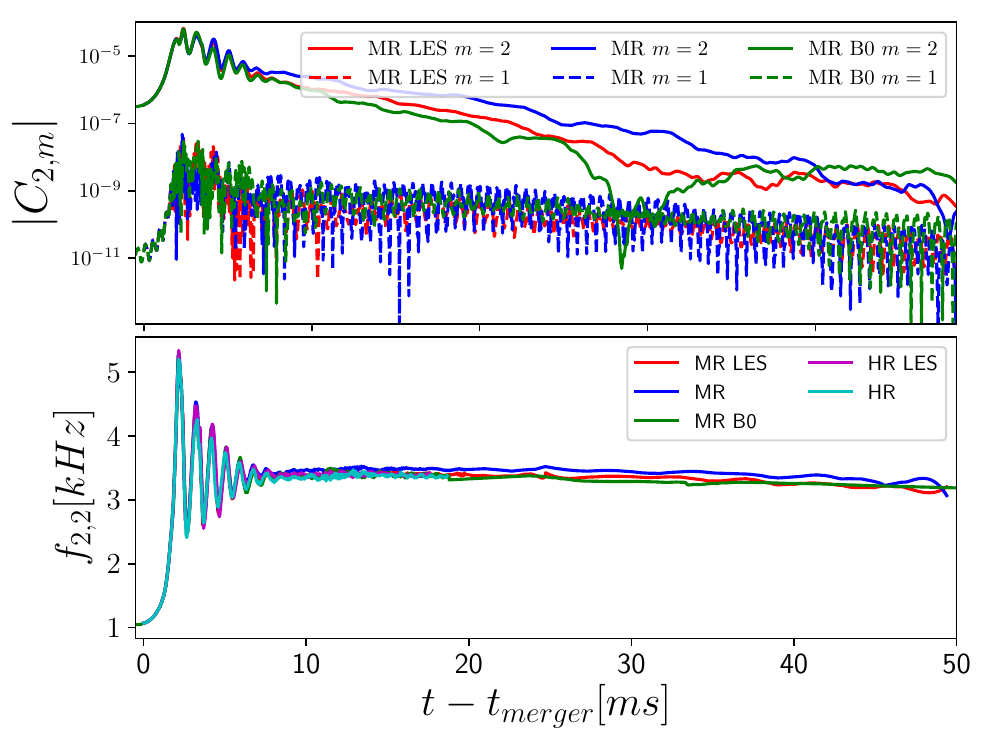}
	\caption{ {\em Evolution of gravitational wave modes}. (Top) The module  $|C_{l=2,m}|$ of the $\Psi_4$ modes $m=1$ and $m=2$, for the medium resolution simulations. The $m=1$ mode is persistent for all the cases (i.e., LES vs no LES, and magnetized vs unmagnetized). (Bottom) Instantaneous frequency of the $l=m=2$ gravitational wave mode for the medium and the high-resolution simulations.
 	 }.
	\label{waveforms_m2}
\end{figure}

The merger of an equal-mass binary leads naturally to a quadrupole (i.e., $m=2$) deformation in the mass distribution of the remnant. As it has been found in previous works (see e.g.~\cite{paschalidis2015one,2016PhRvD..94f4011R,east2016relativistic,2016PhRvD..94d3003L}), a non-axisymmetric instability induces a $m=1$ mode, which develops soon after the merger and at late times it might dominate over the $m=2$ one. The claim in those works is that this $m=1$ mode induces a strong spiral arm in the remnant, which transports angular momentum and contributes to unbind the outer layers of the envelope. One way to analyze the impact of the magnetic fields on the development of this spiral mode is by comparing the $m=1$ mode of the gravitational waves, displayed in the top panel of Figure~\ref{waveforms_m2} for all the medium resolution simulations, either magnetized or unmagnetized. No significant differences are observed in the post-merger signal for the $m=1$ mode, and only moderate ones in the $m=2$ mode. We can conclude that no significant effects are introduced by the magnetic fields on these spiral modes in these timescales, probably because these magnetic fields are not organized yet in large-scale structures.

Finally, one might ask if the presence of strong magnetic fields changes other observable features of the gravitational waves. The amplitudes and  instantaneous frequencies of the GW main mode (i.e., $l=m=2$) are displayed in  Figure~\ref{waveforms_m2}, for all the medium and high resolution simulations. The amplitudes of the unmagnetized and the LES cases matches very well, with a slightly larger deviation of the standard simulation. In the frequencies, only a very small shift is observed in the medium resolution standard simulation with respect to the unmagnetized and to the magnetized LES, confirming that the effect of magnetic fields on these timescales is rather small.

\section{Discussion}\label{sec:discussion}

The combination of high-order numerical schemes, high-resolution and LES techniques with the gradient SGS model, allow us to describe accurately small-scale dynamo effects in magnetized neutron star mergers. 
Our simulations demonstrate that the numerical methods are able to capture the magnetic field amplification within the bulk of the remnant, and the corresponding final saturation level.
These results suggest that LES, combined with the gradient SGS model, can be interpreted as a method to increase the effective resolution of the simulations by at least a factor 2 (and possibly higher). In our scenario, it allowed us to achieve numerical convergence for moderately high resolutions, which would have been otherwise beyond our reach. Let us also notice that, although our numerical setup is not suitable to investigate the convergence in the low density envelope, the results in the bulk are encouraging and we expect benefits also in this regime.

We have considered magnetized neutron stars with realistic magnetic field strength of $10^{11}$ G, an intensity small enough not to alter the topology of the remnant's amplified field. These values represent a tiny fraction of the total energy of the system and are much smaller than commonly used in numerical simulations.  

Our simulations show that, shortly after the merger, the main  magnetic field amplification process is the KHI. A shear layer between the stars, at the time of contact, is prone to unstable modes that develop vortexes at all scales. The RTI might also act in the outer regions of the stars in even shorter timescales, contributing to the development of turbulence. However, it is hard to distinguish such contribution from the rest of the rich dynamics. Then, small-scale dynamo effects driven by these two instabilities amplifies the magnetic field, that is redistributed throughout the remnant by the fluid flow. A turbulent, quasi-stationary state, is achieved after approximately $5$~ms. The average magnetic field strength at the saturation phase is approximately $10^{16}$~G. As expected from an isotropic turbulence, the poloidal and toroidal components of the magnetic field are comparable at saturation. At later times the magnetic field grows linearly due to the winding mechanism, reaching at $t = 50$~ms magnetic energies of $2\times 10^{50}$~ergs (and still growing). Such a magnetic energy value should be considered as a lower bound for the magnetization level reached after BNS mergers with sufficiently long-lived remnant.

This behavior can also be observed in the spectra, which shows also that the magnetic spectrum density almost reaches a equipartition with the kinetic one in the  small scales. The kinetic energy spectra follows the expected Kolmogorov power low $k^{-5/3}$ in the inertial range, while that the typical structure scale of $12$~km.  On the other hand, the magnetic energy spectra displays the expected Kazantsev power law $k^{3/2}$ for large scales, but the typical coherent size is much smaller, $\delta R \sim 700$~m. During the $50$~ms after the merger covered by our longest simulations, the kinetic energy spectra remains mainly unchanged. However, the distribution of the  magnetic energy spectra shifts to larger scales, keeping the Kazantsev power law for small wavenumbers but switching to the Kolmogorov power law for intermediate ones. This change in the distribution is very interesting for several reasons: (i) it indicates the presence of an inverse cascade, with magnetic energy being transferred from small to large scales, (ii) the typical scale of the structures of the magnetic field energy moves from $\delta R \sim 700$~m  to larger scales of $\delta R \sim 2$~km and (iii) the magnetic energy spectra mimics the kinetic one (i.e., not far from a Kolmogorov power law) at small scales. 

These larger magnetic structures can be observed also in the averaged (cylindrical) radial distribution. They show that the toroidal component remains constant during the first $20$~ms, but then it start to grow by a factor of a few in the region near the angular velocity peak. The poloidal component, on the other side, decrease initially during the first $20$~ms, but then started to grow in the inner region $\leq 10$~km. We believe that the winding is responsible for the growth of the toroidal and poloidal magnetic fields, in combination with small-scale dynamo processes.

It is important to stress that we observe no significant redistribution of the differential rotation in the first 50 ms after the merger. This is in agreement with the fact that the MRI is apparently not operating in our simulations. Both things might be explained by the small-scale, randomly oriented structure of the magnetic field during turbulence: the effective Alfven velocity is reduced by a large factor, which depends on the typical sizes $\delta R$ where the mean magnetic field has a well-defined direction and within which the angular momentum is effectively transported. More fundamentally, since the KHI-triggered magnetic field is turbulent and very dynamical, there is no static, large-scale background field over which one can define an unstable perturbation like the MRI. Our expectation is that, as large-scale magnetic fields develop, they can become more efficient on redistributing angular momentum from the bulk to the outer envelope. This could happen via MRI if, at later stages, the magnetic field could acquire a topology clearly dominated by almost stationary large scales in the region where the angular velocity decreases with radius. 

We want to emphasize that, despite our encouraging results about the convergence in the bulk, further studies are needed in order to fully understand the impact of magnetic fields on the remnant's dynamic. Further convergence tests, where the SGS terms are active also in the envelope and where the resolution in the latter is not kept fixed, will be needed to ascertain whether the method can capture the small-scale MHD processes occurring in all of the remnant and follow their development until, ultimately, the possible formation of a magnetically-dominated jet.


\subsection*{Acknowledgments} 
This work was supported by European Union FEDER funds, the Ministry of Science, Innovation and Universities and the Spanish Agencia Estatal de Investigaci\'on grant PID2019-110301GB-I00.
RC acknowledges support from the project PRIN-INAF 2019 ``Short gamma-ray burst jets from binary neutron star mergers''.
JVK kindly acknowledges the CARIPARO Foundation for funding his Ph.D.~fellowship within the Ph.D.~School in Physics at the University of Padova.
DV is funded by the European Research Council (ERC) Starting Grant IMAGINE (grant agreement No. [948582]) under the European Union's Horizon 2020 research and innovation programme. DV's work was also partially supported by the program Unidad de Excelencia María de Maeztu CEX2020-001058-M.
The authors thankfully acknowledge the computer resources at MareNostrum and the technical support provided by Barcelona Supercomputing Center (BSC) through Grant project Long-LESBNS by the $22^{nd}$ PRACE regular call (Proposal 2019215177, P.I. CP and DV).
\clearpage

\bibliographystyle{unsrt}
\bibliography{turbulence}

\begin{thebibliography}{10}

\bibitem{LVC-BNS}
B.~P. {Abbott}, R.~{Abbott}, T.~D. {Abbott}, F.~{Acernese}, K.~{Ackley},
  C.~{Adams}, T.~{Adams}, P.~{Addesso}, R.~X. {Adhikari}, V.~B. {Adya}, and
  et~al.
\newblock {GW170817: Observation of Gravitational Waves from a Binary Neutron
  Star Inspiral}.
\newblock {\em Phys. Rev. Lett.}, 119(16):161101, October 2017.

\bibitem{LVC-MMA}
B.~P. {Abbott}, R.~{Abbott}, T.~D. {Abbott}, F.~{Acernese}, K.~{Ackley},
  C.~{Adams}, T.~{Adams}, P.~{Addesso}, R.~X. {Adhikari}, V.~B. {Adya}, and
  et~al.
\newblock {Multi-messenger Observations of a Binary Neutron Star Merger}.
\newblock {\em Astrophys. J. Lett.}, 848:L12, October 2017.

\bibitem{LVC-GRB}
B.~P. {Abbott}, R.~{Abbott}, T.~D. {Abbott}, F.~{Acernese}, K.~{Ackley},
  C.~{Adams}, T.~{Adams}, P.~{Addesso}, R.~X. {Adhikari}, V.~B. {Adya}, and
  et~al.
\newblock {Gravitational Waves and Gamma-Rays from a Binary Neutron Star
  Merger: GW170817 and GRB 170817A}.
\newblock {\em Astrophys. J. Lett.}, 848:L13, October 2017.

\bibitem{Goldstein2017}
A.~{Goldstein}, P.~{Veres}, E.~{Burns}, M.~S. {Briggs}, R.~{Hamburg},
  D.~{Kocevski}, C.~A. {Wilson-Hodge}, R.~D. {Preece}, S.~{Poolakkil}, O.~J.
  {Roberts}, C.~M. {Hui}, V.~{Connaughton}, J.~{Racusin}, A.~{von Kienlin},
  T.~{Dal Canton}, N.~{Christensen}, T.~{Littenberg}, K.~{Siellez},
  L.~{Blackburn}, J.~{Broida}, E.~{Bissaldi}, W.~H. {Cleveland}, M.~H. {Gibby},
  M.~M. {Giles}, R.~M. {Kippen}, S.~{McBreen}, J.~{McEnery}, C.~A. {Meegan},
  W.~S. {Paciesas}, and M.~{Stanbro}.
\newblock {An Ordinary Short Gamma-Ray Burst with Extraordinary Implications:
  Fermi-GBM Detection of GRB 170817A}.
\newblock {\em \apjl}, 848(2):L14, October 2017.

\bibitem{Savchenko2017}
V.~{Savchenko}, C.~{Ferrigno}, E.~{Kuulkers}, A.~{Bazzano}, E.~{Bozzo},
  S.~{Brandt}, J.~{Chenevez}, T.~J.~L. {Courvoisier}, R.~{Diehl}, A.~{Domingo},
  L.~{Hanlon}, E.~{Jourdain}, A.~{von Kienlin}, P.~{Laurent}, F.~{Lebrun},
  A.~{Lutovinov}, A.~{Martin-Carrillo}, S.~{Mereghetti}, L.~{Natalucci},
  J.~{Rodi}, J.~P. {Roques}, R.~{Sunyaev}, and P.~{Ubertini}.
\newblock {INTEGRAL Detection of the First Prompt Gamma-Ray Signal Coincident
  with the Gravitational-wave Event GW170817}.
\newblock {\em \apjl}, 848(2):L15, October 2017.

\bibitem{Troja2017}
E.~{Troja}, L.~{Piro}, H.~{van Eerten}, R.~T. {Wollaeger}, M.~{Im}, O.~D.
  {Fox}, N.~R. {Butler}, S.~B. {Cenko}, T.~{Sakamoto}, C.~L. {Fryer},
  R.~{Ricci}, A.~{Lien}, R.~E. {Ryan}, O.~{Korobkin}, S.-K. {Lee}, J.~M.
  {Burgess}, W.~H. {Lee}, A.~M. {Watson}, C.~{Choi}, S.~{Covino},
  P.~{D'Avanzo}, C.~J. {Fontes}, J.~B. {Gonz{\'a}lez}, H.~G. {Khandrika},
  J.~{Kim}, S.-L. {Kim}, C.-U. {Lee}, H.~M. {Lee}, A.~{Kutyrev}, G.~{Lim},
  R.~{S{\'a}nchez-Ram{\'{\i}}rez}, S.~{Veilleux}, M.~H. {Wieringa}, and
  Y.~{Yoon}.
\newblock {The X-ray counterpart to the gravitational-wave event GW170817}.
\newblock {\em Nature}, 551:71--74, November 2017.

\bibitem{Margutti2017}
R.~{Margutti}, E.~{Berger}, W.~{Fong}, C.~{Guidorzi}, K.~D. {Alexander}, B.~D.
  {Metzger}, P.~K. {Blanchard}, P.~S. {Cowperthwaite}, R.~{Chornock},
  T.~{Eftekhari}, M.~{Nicholl}, V.~A. {Villar}, P.~K.~G. {Williams},
  J.~{Annis}, D.~A. {Brown}, H.~{Chen}, Z.~{Doctor}, J.~A. {Frieman}, D.~E.
  {Holz}, M.~{Sako}, and M.~{Soares-Santos}.
\newblock {The Electromagnetic Counterpart of the Binary Neutron Star Merger
  LIGO/Virgo GW170817. V. Rising X-Ray Emission from an Off-axis Jet}.
\newblock {\em Astrophys. J. Lett.}, 848:L20, October 2017.

\bibitem{Hallinan2017}
G.~{Hallinan}, A.~{Corsi}, K.~P. {Mooley}, K.~{Hotokezaka}, E.~{Nakar}, M.~M.
  {Kasliwal}, D.~L. {Kaplan}, D.~A. {Frail}, S.~T. {Myers}, T.~{Murphy},
  K.~{De}, D.~{Dobie}, J.~R. {Allison}, K.~W. {Bannister}, V.~{Bhalerao},
  P.~{Chandra}, T.~E. {Clarke}, S.~{Giacintucci}, A.~Y.~Q. {Ho}, A.~{Horesh},
  N.~E. {Kassim}, S.~R. {Kulkarni}, E.~{Lenc}, F.~J. {Lockman}, C.~{Lynch},
  D.~{Nichols}, S.~{Nissanke}, N.~{Palliyaguru}, W.~M. {Peters}, T.~{Piran},
  J.~{Rana}, E.~M. {Sadler}, and L.~P. {Singer}.
\newblock {A radio counterpart to a neutron star merger}.
\newblock {\em Science}, 358:1579--1583, December 2017.

\bibitem{Alexander2017}
K.~D. {Alexander}, E.~{Berger}, W.~{Fong}, P.~K.~G. {Williams}, C.~{Guidorzi},
  R.~{Margutti}, B.~D. {Metzger}, J.~{Annis}, P.~K. {Blanchard}, D.~{Brout},
  D.~A. {Brown}, H.-Y. {Chen}, R.~{Chornock}, P.~S. {Cowperthwaite},
  M.~{Drout}, T.~{Eftekhari}, J.~{Frieman}, D.~E. {Holz}, M.~{Nicholl},
  A.~{Rest}, M.~{Sako}, M.~{Soares-Santos}, and V.~A. {Villar}.
\newblock {The Electromagnetic Counterpart of the Binary Neutron Star Merger
  LIGO/Virgo GW170817. VI. Radio Constraints on a Relativistic Jet and
  Predictions for Late-time Emission from the Kilonova Ejecta}.
\newblock {\em Astrophys. J. Lett.}, 848:L21, October 2017.

\bibitem{Mooley2018a}
K.~P. {Mooley}, E.~{Nakar}, K.~{Hotokezaka}, G.~{Hallinan}, A.~{Corsi}, D.~A.
  {Frail}, A.~{Horesh}, T.~{Murphy}, E.~{Lenc}, D.~L. {Kaplan}, K.~{de},
  D.~{Dobie}, P.~{Chandra}, A.~{Deller}, O.~{Gottlieb}, M.~M. {Kasliwal}, S.~R.
  {Kulkarni}, S.~T. {Myers}, S.~{Nissanke}, T.~{Piran}, C.~{Lynch},
  V.~{Bhalerao}, S.~{Bourke}, K.~W. {Bannister}, and L.~P. {Singer}.
\newblock {A mildly relativistic wide-angle outflow in the neutron-star merger
  event GW170817}.
\newblock {\em Nature}, 554:207--210, February 2018.

\bibitem{Lazzati2018}
D.~{Lazzati}, R.~{Perna}, B.~J. {Morsony}, D.~{Lopez-Camara}, M.~{Cantiello},
  R.~{Ciolfi}, B.~{Giacomazzo}, and J.~C. {Workman}.
\newblock {Late Time Afterglow Observations Reveal a Collimated Relativistic
  Jet in the Ejecta of the Binary Neutron Star Merger GW170817}.
\newblock {\em Phys. Rev. Lett.}, 120(24):241103, June 2018.

\bibitem{Lyman2018}
J.~D. {Lyman}, G.~P. {Lamb}, A.~J. {Levan}, I.~{Mandel}, N.~R. {Tanvir},
  S.~{Kobayashi}, B.~{Gompertz}, J.~{Hjorth}, A.~S. {Fruchter}, T.~{Kangas},
  D.~{Steeghs}, I.~A. {Steele}, Z.~{Cano}, C.~{Copperwheat}, P.~A. {Evans},
  J.~P.~U. {Fynbo}, C.~{Gall}, M.~{Im}, L.~{Izzo}, P.~{Jakobsson},
  B.~{Milvang-Jensen}, P.~{O'Brien}, J.~P. {Osborne}, E.~{Palazzi}, D.~A.
  {Perley}, E.~{Pian}, S.~{Rosswog}, A.~{Rowlinson}, S.~{Schulze}, E.~R.
  {Stanway}, P.~{Sutton}, C.~C. {Th{\"o}ne}, A.~{de Ugarte Postigo}, D.~J.
  {Watson}, K.~{Wiersema}, and R.~A.~M.~J. {Wijers}.
\newblock {The optical afterglow of the short gamma-ray burst associated with
  GW170817}.
\newblock {\em Nature Astr.}, 2:751--754, July 2018.

\bibitem{Alexander2018}
K.~D. {Alexander}, R.~{Margutti}, P.~K. {Blanchard}, W.~{Fong}, E.~{Berger},
  A.~{Hajela}, T.~{Eftekhari}, R.~{Chornock}, P.~S. {Cowperthwaite},
  D.~{Giannios}, C.~{Guidorzi}, A.~{Kathirgamaraju}, A.~{MacFadyen}, B.~D.
  {Metzger}, M.~{Nicholl}, L.~{Sironi}, V.~A. {Villar}, P.~K.~G. {Williams},
  X.~{Xie}, and J.~{Zrake}.
\newblock {A Decline in the X-Ray through Radio Emission from GW170817
  Continues to Support an Off-axis Structured Jet}.
\newblock {\em Astrophys. J. Lett.}, 863:L18, August 2018.

\bibitem{Mooley2018b}
K.~P. {Mooley}, A.~T. {Deller}, O.~{Gottlieb}, E.~{Nakar}, G.~{Hallinan},
  S.~{Bourke}, D.~A. {Frail}, A.~{Horesh}, A.~{Corsi}, and K.~{Hotokezaka}.
\newblock {Superluminal motion of a relativistic jet in the neutron-star merger
  GW170817}.
\newblock {\em Nature}, 561:355--359, September 2018.

\bibitem{Ghirlanda2019}
G.~{Ghirlanda}, O.~S. {Salafia}, Z.~{Paragi}, M.~{Giroletti}, J.~{Yang},
  B.~{Marcote}, J.~{Blanchard}, I.~{Agudo}, T.~{An}, M.~G. {Bernardini},
  R.~{Beswick}, M.~{Branchesi}, S.~{Campana}, C.~{Casadio}, E.~{Chassand
  e-Mottin}, M.~{Colpi}, S.~{Covino}, P.~{D'Avanzo}, V.~{D'Elia}, S.~{Frey},
  M.~{Gawronski}, G.~{Ghisellini}, L.~I. {Gurvits}, P.~G. {Jonker}, H.~J. {van
  Langevelde}, A.~{Melandri}, J.~{Moldon}, L.~{Nava}, A.~{Perego}, M.~A.
  {Perez-Torres}, C.~{Reynolds}, R.~{Salvaterra}, G.~{Tagliaferri},
  T.~{Venturi}, S.~D. {Vergani}, and M.~{Zhang}.
\newblock {Compact radio emission indicates a structured jet was produced by a
  binary neutron star merger}.
\newblock {\em Science}, 363(6430):968--971, Mar 2019.

\bibitem{Arcavi2017}
Iair {Arcavi}, Griffin {Hosseinzadeh}, D.~Andrew {Howell}, Curtis {McCully},
  Dovi {Poznanski}, Daniel {Kasen}, Jennifer {Barnes}, Michael {Zaltzman},
  Sergiy {Vasylyev}, Dan {Maoz}, and Stefano {Valenti}.
\newblock {Optical emission from a kilonova following a
  gravitational-wave-detected neutron-star merger}.
\newblock {\em Nature}, 551(7678):64--66, Nov 2017.

\bibitem{Coulter2017}
D.~A. {Coulter}, R.~J. {Foley}, C.~D. {Kilpatrick}, M.~R. {Drout}, A.~L.
  {Piro}, B.~J. {Shappee}, M.~R. {Siebert}, J.~D. {Simon}, N.~{Ulloa},
  D.~{Kasen}, B.~F. {Madore}, A.~{Murguia-Berthier}, Y.~C. {Pan}, J.~X.
  {Prochaska}, E.~{Ramirez-Ruiz}, A.~{Rest}, and C.~{Rojas-Bravo}.
\newblock {Swope Supernova Survey 2017a (SSS17a), the optical counterpart to a
  gravitational wave source}.
\newblock {\em Science}, 358(6370):1556--1558, Dec 2017.

\bibitem{Pian2017}
E.~{Pian}, P.~{D'Avanzo}, S.~{Benetti}, M.~{Branchesi}, E.~{Brocato},
  S.~{Campana}, E.~{Cappellaro}, S.~{Covino}, V.~{D'Elia}, J.~P.~U. {Fynbo},
  F.~{Getman}, G.~{Ghirland a}, G.~{Ghisellini}, A.~{Grado}, G.~{Greco},
  J.~{Hjorth}, C.~{Kouveliotou}, A.~{Levan}, L.~{Limatola}, D.~{Malesani},
  P.~A. {Mazzali}, A.~{Melandri}, P.~{M{\o}ller}, L.~{Nicastro}, E.~{Palazzi},
  S.~{Piranomonte}, A.~{Rossi}, O.~S. {Salafia}, J.~{Selsing}, G.~{Stratta},
  M.~{Tanaka}, N.~R. {Tanvir}, L.~{Tomasella}, D.~{Watson}, S.~{Yang},
  L.~{Amati}, L.~A. {Antonelli}, S.~{Ascenzi}, M.~G. {Bernardini},
  M.~{Bo{\"e}r}, F.~{Bufano}, A.~{Bulgarelli}, M.~{Capaccioli}, P.~{Casella},
  A.~J. {Castro-Tirado}, E.~{Chassande-Mottin}, R.~{Ciolfi}, C.~M.
  {Copperwheat}, M.~{Dadina}, G.~{De Cesare}, A.~{di Paola}, Y.~Z. {Fan},
  B.~{Gendre}, G.~{Giuffrida}, A.~{Giunta}, L.~K. {Hunt}, G.~L. {Israel}, Z.~P.
  {Jin}, M.~M. {Kasliwal}, S.~{Klose}, M.~{Lisi}, F.~{Longo}, E.~{Maiorano},
  M.~{Mapelli}, N.~{Masetti}, L.~{Nava}, B.~{Patricelli}, D.~{Perley},
  A.~{Pescalli}, T.~{Piran}, A.~{Possenti}, L.~{Pulone}, M.~{Razzano},
  R.~{Salvaterra}, P.~{Schipani}, M.~{Spera}, A.~{Stamerra}, L.~{Stella},
  G.~{Tagliaferri}, V.~{Testa}, E.~{Troja}, M.~{Turatto}, S.~D. {Vergani}, and
  D.~{Vergani}.
\newblock {Spectroscopic identification of r-process nucleosynthesis in a
  double neutron-star merger}.
\newblock {\em Nature}, 551(7678):67--70, Nov 2017.

\bibitem{Smartt2017}
S.~J. {Smartt}, T.~W. {Chen}, A.~{Jerkstrand}, M.~{Coughlin}, E.~{Kankare},
  S.~A. {Sim}, M.~{Fraser}, C.~{Inserra}, K.~{Maguire}, K.~C. {Chambers}, M.~E.
  {Huber}, T.~{Kr{\"u}hler}, G.~{Leloudas}, M.~{Magee}, L.~J. {Shingles}, K.~W.
  {Smith}, D.~R. {Young}, J.~{Tonry}, R.~{Kotak}, A.~{Gal-Yam}, J.~D. {Lyman},
  D.~S. {Homan}, C.~{Agliozzo}, J.~P. {Anderson}, C.~R. {Angus}, C.~{Ashall},
  C.~{Barbarino}, F.~E. {Bauer}, M.~{Berton}, M.~T. {Botticella}, M.~{Bulla},
  J.~{Bulger}, G.~{Cannizzaro}, Z.~{Cano}, R.~{Cartier}, A.~{Cikota},
  P.~{Clark}, A.~{De Cia}, M.~{Della Valle}, L.~{Denneau}, M.~{Dennefeld},
  L.~{Dessart}, G.~{Dimitriadis}, N.~{Elias-Rosa}, R.~E. {Firth},
  H.~{Flewelling}, A.~{Fl{\"o}rs}, A.~{Franckowiak}, C.~{Frohmaier},
  L.~{Galbany}, S.~{Gonz{\'a}lez-Gait{\'a}n}, J.~{Greiner}, M.~{Gromadzki},
  A.~Nicuesa {Guelbenzu}, C.~P. {Guti{\'e}rrez}, A.~{Hamanowicz}, L.~{Hanlon},
  J.~{Harmanen}, K.~E. {Heintz}, A.~{Heinze}, M.~S. {Hernandez}, S.~T.
  {Hodgkin}, I.~M. {Hook}, L.~{Izzo}, P.~A. {James}, P.~G. {Jonker}, W.~E.
  {Kerzendorf}, S.~{Klose}, Z.~{Kostrzewa-Rutkowska}, M.~{Kowalski},
  M.~{Kromer}, H.~{Kuncarayakti}, A.~{Lawrence}, T.~B. {Lowe}, E.~A. {Magnier},
  I.~{Manulis}, A.~{Martin-Carrillo}, S.~{Mattila}, O.~{McBrien},
  A.~{M{\"u}ller}, J.~{Nordin}, D.~{O'Neill}, F.~{Onori}, J.~T. {Palmerio},
  A.~{Pastorello}, F.~{Patat}, G.~{Pignata}, Ph. {Podsiadlowski}, M.~L. {Pumo},
  S.~J. {Prentice}, A.~{Rau}, A.~{Razza}, A.~{Rest}, T.~{Reynolds}, R.~{Roy},
  A.~J. {Ruiter}, K.~A. {Rybicki}, L.~{Salmon}, P.~{Schady}, A.~S.~B.
  {Schultz}, T.~{Schweyer}, I.~R. {Seitenzahl}, M.~{Smith}, J.~{Sollerman},
  B.~{Stalder}, C.~W. {Stubbs}, M.~{Sullivan}, H.~{Szegedi}, F.~{Taddia},
  S.~{Taubenberger}, G.~{Terreran}, B.~{van Soelen}, J.~{Vos}, R.~J.
  {Wainscoat}, N.~A. {Walton}, C.~{Waters}, H.~{Weiland}, M.~{Willman},
  P.~{Wiseman}, D.~E. {Wright}, {\L}.~{Wyrzykowski}, and O.~{Yaron}.
\newblock {A kilonova as the electromagnetic counterpart to a
  gravitational-wave source}.
\newblock {\em Nature}, 551(7678):75--79, Nov 2017.

\bibitem{Kasen2017}
Daniel {Kasen}, Brian {Metzger}, Jennifer {Barnes}, Eliot {Quataert}, and
  Enrico {Ramirez-Ruiz}.
\newblock {Origin of the heavy elements in binary neutron-star mergers from a
  gravitational-wave event}.
\newblock {\em Nature}, 551(7678):80--84, Nov 2017.

\bibitem{Metzger2019LRR}
Brian~D. {Metzger}.
\newblock {Kilonovae}.
\newblock {\em Liv. Rev. Rel.}, 23(1):1, Dec 2019.

\bibitem{LVC-170817properties}
B.~P. {Abbott}, R.~{Abbott}, T.~D. {Abbott}, F.~{Acernese}, K.~{Ackley},
  C.~{Adams}, T.~{Adams}, P.~{Addesso}, R.~X. {Adhikari}, V.~B. {Adya}, and
  et~al.
\newblock {Properties of the Binary Neutron Star Merger GW170817}.
\newblock {\em Phys. Rev. X}, 9(1):011001, Jan 2019.

\bibitem{LVC-Hubble}
B.~P. {Abbott}, R.~{Abbott}, T.~D. {Abbott}, F.~{Acernese}, K.~{Ackley},
  C.~{Adams}, T.~{Adams}, P.~{Addesso}, R.~X. {Adhikari}, V.~B. {Adya}, and
  et~al.
\newblock {A gravitational-wave standard siren measurement of the Hubble
  constant}.
\newblock {\em Nature}, 551:85--88, November 2017.

\bibitem{Ciolfi2020c}
Riccardo {Ciolfi}.
\newblock {Binary neutron star mergers after GW170817}.
\newblock {\em Front. Astron. Sp. Sci.}, 7:27, June 2020.

\bibitem{Paschalidis2017}
Vasileios {Paschalidis}.
\newblock {General relativistic simulations of compact binary mergers as
  engines for short gamma-ray bursts}.
\newblock {\em Classical and Quantum Gravity}, 34(8):084002, April 2017.

\bibitem{Duez2019}
Matthew~D. {Duez} and Yosef {Zlochower}.
\newblock {Numerical relativity of compact binaries in the 21st century}.
\newblock {\em Reports on Progress in Physics}, 82(1):016902, January 2019.

\bibitem{Shibata2019}
Masaru {Shibata} and Kenta {Hotokezaka}.
\newblock {Merger and Mass Ejection of Neutron Star Binaries}.
\newblock {\em Annual Review of Nuclear and Particle Science}, 69:41--64,
  October 2019.

\bibitem{Ciolfi2020b}
Riccardo {Ciolfi}.
\newblock {The key role of magnetic fields in binary neutron star mergers}.
\newblock {\em Gen. Rel. Grav.}, 52(6):59, June 2020.

\bibitem{Palenzuela2020}
Carlos {Palenzuela}.
\newblock {Introduction to Numerical Relativity}.
\newblock {\em Frontiers in Astronomy and Space Sciences}, 7:58, September
  2020.

\bibitem{price06}
D.~J. {Price} and S.~{Rosswog}.
\newblock {Producing Ultrastrong Magnetic Fields in Neutron Star Mergers}.
\newblock {\em Science}, 312:719--722, May 2006.

\bibitem{kiuchi15}
K.~{Kiuchi}, P.~{Cerd{\'a}-Dur{\'a}n}, K.~{Kyutoku}, Y.~{Sekiguchi}, and
  M.~{Shibata}.
\newblock {Efficient magnetic-field amplification due to the Kelvin-Helmholtz
  instability in binary neutron star mergers}.
\newblock {\em \prd}, 92(12):124034, December 2015.

\bibitem{2021ApJ...921...75S}
V.~{Skoutnev}, E.~R. {Most}, A.~{Bhattacharjee}, and A.~A. {Philippov}.
\newblock {Scaling of Small-scale Dynamo Properties in the Rayleigh-Taylor
  Instability}.
\newblock {\em \apj}, 921(1):75, November 2021.

\bibitem{balbus91}
S.~A. {Balbus} and J.~F. {Hawley}.
\newblock {A powerful local shear instability in weakly magnetized disks. I -
  Linear analysis. II - Nonlinear evolution}.
\newblock {\em \apj}, 376:214--233, July 1991.

\bibitem{balbus98}
S.~A. {Balbus} and J.~F. {Hawley}.
\newblock {Instability, turbulence, and enhanced transport in accretion disks}.
\newblock {\em Reviews of Modern Physics}, 70:1--53, January 1998.

\bibitem{PhysRevLett.96.031101}
Matthew~D. Duez, Yuk~Tung Liu, Stuart~L. Shapiro, Masaru Shibata, and
  Branson~C. Stephens.
\newblock Collapse of magnetized hypermassive neutron stars in general
  relativity.
\newblock {\em Phys. Rev. Lett.}, 96:031101, Jan 2006.

\bibitem{2013PhRvD..87l1302S}
Daniel~M. {Siegel}, Riccardo {Ciolfi}, Abraham~I. {Harte}, and Luciano
  {Rezzolla}.
\newblock {Magnetorotational instability in relativistic hypermassive neutron
  stars}.
\newblock {\em \prd}, 87(12):121302, June 2013.

\bibitem{kiuchi14}
K.~{Kiuchi}, K.~{Kyutoku}, Y.~{Sekiguchi}, M.~{Shibata}, and T.~{Wada}.
\newblock {High resolution numerical relativity simulations for the merger of
  binary magnetized neutron stars}.
\newblock {\em \prd}, 90(4):041502, August 2014.

\bibitem{kiuchi18}
K.~{Kiuchi}, K.~{Kyutoku}, Y.~{Sekiguchi}, and M.~{Shibata}.
\newblock {Global simulations of strongly magnetized remnant massive neutron
  stars formed in binary neutron star mergers}.
\newblock {\em \prd}, 97(12):124039, June 2018.

\bibitem{ruiz16}
M.~{Ruiz}, R.~N. {Lang}, V.~{Paschalidis}, and S.~L. {Shapiro}.
\newblock {Binary Neutron Star Mergers: A Jet Engine for Short Gamma-Ray
  Bursts}.
\newblock {\em \apjl}, 824:L6, June 2016.

\bibitem{ciolfi2019}
Riccardo Ciolfi, Wolfgang Kastaun, Jay~Vijay Kalinani, and Bruno Giacomazzo.
\newblock First 100 ms of a long-lived magnetized neutron star formed in a
  binary neutron star merger.
\newblock {\em Physical Review D}, 100(2):023005, 2019.

\bibitem{ciolfi2020collimated}
Riccardo Ciolfi.
\newblock Collimated outflows from long-lived binary neutron star merger
  remnants.
\newblock {\em Monthly Notices of the Royal Astronomical Society: Letters},
  495(1):L66--L70, 2020.

\bibitem{ruiz2020}
Milton Ruiz, Antonios Tsokaros, and Stuart~L Shapiro.
\newblock Magnetohydrodynamic simulations of binary neutron star mergers in
  general relativity: Effects of magnetic field orientation on jet launching.
\newblock {\em Physical Review D}, 101(6):064042, 2020.

\bibitem{mosta2020}
Philipp M{\"o}sta, David Radice, Roland Haas, Erik Schnetter, and Sebastiano
  Bernuzzi.
\newblock A magnetar engine for short grbs and kilonovae.
\newblock {\em arXiv preprint arXiv:2003.06043}, 2020.

\bibitem{tauris17}
T.~M. {Tauris}, M.~{Kramer}, P.~C.~C. {Freire}, N.~{Wex}, H.~T. {Janka},
  N.~{Langer}, Ph. {Podsiadlowski}, E.~{Bozzo}, S.~{Chaty}, M.~U. {Kruckow},
  E.~P.~J. {van den Heuvel}, J.~{Antoniadis}, R.~P. {Breton}, and D.~J.
  {Champion}.
\newblock {Formation of Double Neutron Star Systems}.
\newblock {\em \apj}, 846(2):170, Sep 2017.

\bibitem{zhiyin15}
Zhiyin Yang.
\newblock Large-eddy simulation: Past, present and the future.
\newblock {\em Chinese Journal of Aeronautics}, 91, 12 2014.

\bibitem{bucciantini13}
N.~{Bucciantini} and L.~{Del Zanna}.
\newblock {A fully covariant mean-field dynamo closure for numerical 3 + 1
  resistive GRMHD}.
\newblock {\em \mnras}, 428:71--85, January 2013.

\bibitem{giacomazzo15}
B.~{Giacomazzo}, J.~{Zrake}, P.~C. {Duffell}, A.~I. {MacFadyen}, and
  R.~{Perna}.
\newblock {Producing Magnetar Magnetic Fields in the Merger of Binary Neutron
  Stars}.
\newblock {\em \apj}, 809:39, August 2015.

\bibitem{palenzuela15}
C.~{Palenzuela}, S.~L. {Liebling}, D.~{Neilsen}, L.~{Lehner}, O.~L.
  {Caballero}, E.~{O'Connor}, and M.~{Anderson}.
\newblock {Effects of the microphysical equation of state in the mergers of
  magnetized neutron stars with neutrino cooling}.
\newblock {\em \prd}, 92(4):044045, August 2015.

\bibitem{duez2004}
Matthew~D Duez, Yuk~Tung Liu, Stuart~L Shapiro, and Branson~C Stephens.
\newblock General relativistic hydrodynamics with viscosity: Contraction,
  catastrophic collapse, and disk formation in hypermassive neutron stars.
\newblock {\em Physical Review D}, 69(10):104030, 2004.

\bibitem{shibata2017general}
Masaru Shibata, Kenta Kiuchi, and Yu-ichiro Sekiguchi.
\newblock General relativistic viscous hydrodynamics of differentially rotating
  neutron stars.
\newblock {\em Physical Review D}, 95(8):083005, 2017.

\bibitem{radice17}
D.~{Radice}.
\newblock {General-relativistic Large-eddy Simulations of Binary Neutron Star
  Mergers}.
\newblock {\em \apjl}, 838:L2, March 2017.

\bibitem{fujibayashi2020}
Sho Fujibayashi, Masaru Shibata, Shinya Wanajo, Kenta Kiuchi, Koutarou Kyutoku,
  and Yuichiro Sekiguchi.
\newblock Mass ejection from disks surrounding a low-mass black hole: Viscous
  neutrino-radiation hydrodynamics simulation in full general relativity.
\newblock {\em Physical Review D}, 101(8):083029, 2020.

\bibitem{radice2020}
David Radice.
\newblock Binary neutron star merger simulations with a calibrated turbulence
  model.
\newblock {\em arXiv preprint arXiv:2005.09002}, 2020.

\bibitem{leonard75}
A.~{Leonard}.
\newblock {Energy Cascade in Large-Eddy Simulations of Turbulent Fluid Flows}.
\newblock {\em Advances in Geophysics}, 18:237--248, 1975.

\bibitem{muller02a}
W.-C. {M{\"u}ller} and D.~{Carati}.
\newblock {Dynamic gradient-diffusion subgrid models for incompressible
  magnetohydrodynamic turbulence}.
\newblock {\em Physics of Plasmas}, 9:824--834, March 2002.

\bibitem{vigano19b}
Daniele {Vigan{\`o}}, Ricard {Aguilera-Miret}, and Carlos {Palenzuela}.
\newblock {Extension of the subgrid-scale gradient model for compressible
  magnetohydrodynamics turbulent instabilities}.
\newblock {\em Physics of Fluids}, 31(10):105102, Oct 2019.

\bibitem{carrasco19}
Federico Carrasco, Daniele Vigan{\`o}, and Carlos Palenzuela.
\newblock Gradient subgrid-scale model for relativistic mhd large-eddy
  simulations.
\newblock {\em Physical Review D}, 101(6):063003, 2020.

\bibitem{vigano20}
Daniele {Vigan{\`o}}, Ricard {Aguilera-Miret}, Federico {Carrasco}, Borja
  {Mi{\~n}ano}, and Carlos {Palenzuela}.
\newblock {General relativistic MHD large eddy simulations with gradient
  subgrid-scale model}.
\newblock {\em \prd}, 101(12):123019, June 2020.

\bibitem{aguilera2020}
Ricard Aguilera-Miret, Daniele Vigan{\`o}, Federico Carrasco, Borja Mi{\~n}ano,
  and Carlos Palenzuela.
\newblock Turbulent magnetic-field amplification in the first 10 milliseconds
  after a binary neutron star merger: Comparing high-resolution and large-eddy
  simulations.
\newblock {\em Physical Review D}, 102(10):103006, 2020.

\bibitem{read09}
J.~S. {Read}, B.~D. {Lackey}, B.~J. {Owen}, and J.~L. {Friedman}.
\newblock Constraints on a phenomenologically parametrized neutron-star
  equation of state.
\newblock {\em Physical Review D}, 79(12), Jun 2009.

\bibitem{Endrizzi2016}
A~Endrizzi, R~Ciolfi, B~Giacomazzo, W~Kastaun, and T~Kawamura.
\newblock General relativistic magnetohydrodynamic simulations of binary
  neutron star mergers with the apr4 equation of state.
\newblock {\em Class. Quantum Grav.}, 33(16):164001, 2016.

\bibitem{bonabook}
C.~{Bona}, C.~{Palenzuela-Luque}, and C.~{Bona-Casas}, editors.
\newblock {\em {Elements of Numerical Relativity and Relativistic
  Hydrodynamics}}, volume 783 of {\em Lecture Notes in Physics, Berlin Springer
  Verlag}, 2009.

\bibitem{alic12}
Daniela {Alic}, Carles {Bona-Casas}, Carles {Bona}, Luciano {Rezzolla}, and
  Carlos {Palenzuela}.
\newblock {Conformal and covariant formulation of the Z4 system with
  constraint-violation damping}.
\newblock {\em \prd}, 85(6):064040, Mar 2012.

\bibitem{bezares17}
Miguel Bezares, Carlos Palenzuela, and Carles Bona.
\newblock Final fate of compact boson star mergers.
\newblock {\em Phys. Rev. D}, 95:124005, Jun 2017.

\bibitem{palenzuela18}
C.~{Palenzuela}, B.~{Mi{\~n}ano}, D.~{Vigan{\`o}}, A.~{Arbona},
  C.~{Bona-Casas}, A.~{Rigo}, M.~{Bezares}, C.~{Bona}, and J.~{Mass{\'o}}.
\newblock {A Simflowny-based finite-difference code for high-performance
  computing in numerical relativity}.
\newblock {\em Classical and Quantum Gravity}, 35(18):185007, September 2018.

\bibitem{arbona13}
A.~{Arbona}, A.~{Artigues}, C.~{Bona-Casas}, J.~{Mass{\'o}}, B.~{Mi{\~n}ano},
  A.~{Rigo}, M.~{Trias}, and C.~{Bona}.
\newblock {Simflowny: A general-purpose platform for the management of physical
  models and simulation problems}.
\newblock {\em Computer Physics Communications}, 184:2321--2331, October 2013.

\bibitem{arbona18}
A.~{Arbona}, B.~{Mi{\~n}ano}, A.~{Rigo}, C.~{Bona}, C.~{Palenzuela},
  A.~{Artigues}, C.~{Bona-Casas}, and J.~{Mass{\'o}}.
\newblock {Simflowny 2: An upgraded platform for scientific modelling and
  simulation}.
\newblock {\em Computer Physics Communications}, 229:170--181, August 2018.

\bibitem{hornung02}
Richard~D. Hornung and Scott~R. Kohn.
\newblock Managing application complexity in the samrai object-oriented
  framework.
\newblock {\em Concurrency and Computation: Practice and Experience},
  14(5):347--368, 2002.

\bibitem{gunney16}
Brian~T.N. Gunney and Robert~W. Anderson.
\newblock Advances in patch-based adaptive mesh refinement scalability.
\newblock {\em Journal of Parallel and Distributed Computing}, 89:65 -- 84,
  2016.

\bibitem{vigano19}
Daniele {Vigan{\`o}}, David {Mart{\'\i}nez-G{\'o}mez}, Jos{\'e}~A. {Pons},
  Carlos {Palenzuela}, Federico {Carrasco}, Borja {Mi{\~n}ano}, Antoni
  {Arbona}, Carles {Bona}, and Joan {Mass{\'o}}.
\newblock {A Simflowny-based high-performance 3D code for the generalized
  induction equation}.
\newblock {\em Computer Physics Communications}, 237:168--183, Apr 2019.

\bibitem{liebling20}
S.~L. {Liebling}, C.~{Palenzuela}, and L.~{Lehner}.
\newblock Toward fidelity and scalability in non-vacuum mergers.
\newblock {\em Classical and Quantum Gravity}, 37(13):135006, jun 2020.

\bibitem{shu98}
Chi-Wang Shu.
\newblock {\em Essentially non-oscillatory and weighted essentially
  non-oscillatory schemes for hyperbolic conservation laws}, pages 325--432.
\newblock Springer Berlin Heidelberg, Berlin, Heidelberg, 1998.

\bibitem{suresh97}
A.~Suresh and H.T. Huynh.
\newblock Accurate monotonicity-preserving schemes with runge–kutta time
  stepping.
\newblock {\em Journal of Computational Physics}, 136(1):83 -- 99, 1997.

\bibitem{McCorquodale:2011}
Peter McCorquodale and Phillip Colella.
\newblock A high-order finite-volume method for conservation laws on locally
  refined grids.
\newblock {\em Commun. Appl. Math. Comput. Sci.}, 6(1):1--25, 2011.

\bibitem{Mongwane:2015}
B.~{Mongwane}.
\newblock {Toward a consistent framework for high order mesh refinement schemes
  in numerical relativity}.
\newblock {\em General Relativity and Gravitation}, 47:60, May 2015.

\bibitem{kastaun20}
Wolfgang {Kastaun}, Jay {Vijay Kalinani}, and Riccardo {Ciolfi}.
\newblock {Robust Recovery of Primitive Variables in Relativistic Ideal
  Magnetohydrodynamics}.
\newblock {\em arXiv e-prints}, page arXiv:2005.01821, May 2020.

\bibitem{lorene}
{\sc Lorene} home page.
\newblock \url{http://www.lorene.obspm.fr/}, 2010.

\bibitem{2014ApJS..212....6O}
S.~A. {Olausen} and V.~M. {Kaspi}.
\newblock {The McGill Magnetar Catalog}.
\newblock {\em \apjs}, 212(1):6, May 2014.

\bibitem{aguilera21}
Ricard Aguilera-Miret, Daniele Vigan{\`{o}}, and Carlos Palenzuela.
\newblock Universality of the turbulent magnetic field in hypermassive neutron
  stars produced by binary mergers.
\newblock {\em The Astrophysical Journal Letters}, 926(2):L31, feb 2022.

\bibitem{rezbish}
N.~T. {Bishop} and L.~{Rezzolla}.
\newblock {Extraction of gravitational waves in numerical relativity}.
\newblock {\em Living Reviews in Relativity}, 19:2, October 2016.

\bibitem{brugman}
B.~{Br{\"u}gmann}, J.~A. {Gonz{\'a}lez}, M.~{Hannam}, S.~{Husa}, U.~{Sperhake},
  and W.~{Tichy}.
\newblock {Calibration of moving puncture simulations}.
\newblock {\em \prd}, 77(2):024027, January 2008.

\bibitem{shibatabook}
Masaru {Shibata}.
\newblock {\em {Numerical Relativity}}.
\newblock 2016.

\bibitem{kastaun21}
W.~Kastaun and F.~Ohme.
\newblock Numerical inside view of hypermassive remnant models for gw170817.
\newblock {\em Phys. Rev. D}, 104:023001, Jul 2021.

\bibitem{CiolfiKalinani20}
Riccardo {Ciolfi} and Jay~Vijay {Kalinani}.
\newblock {Magnetically Driven Baryon Winds from Binary Neutron Star Merger
  Remnants and the Blue Kilonova of 2017 August}.
\newblock {\em Astrophys. J. Lett.}, 900(2):L35, September 2020.

\bibitem{paschalidis2015one}
Vasileios Paschalidis, William~E East, Frans Pretorius, and Stuart~L Shapiro.
\newblock One-arm spiral instability in hypermassive neutron stars formed by
  dynamical-capture binary neutron star mergers.
\newblock {\em Physical Review D}, 92(12):121502, 2015.

\bibitem{2016PhRvD..94f4011R}
David {Radice}, Sebastiano {Bernuzzi}, and Christian~D. {Ott}.
\newblock {One-armed spiral instability in neutron star mergers and its
  detectability in gravitational waves}.
\newblock {\em \prd}, 94(6):064011, September 2016.

\bibitem{east2016relativistic}
William~E East, Vasileios Paschalidis, Frans Pretorius, and Stuart~L Shapiro.
\newblock Relativistic simulations of eccentric binary neutron star mergers:
  One-arm spiral instability and effects of neutron star spin.
\newblock {\em Physical Review D}, 93(2):024011, 2016.

\bibitem{2016PhRvD..94d3003L}
Luis {Lehner}, Steven~L. {Liebling}, Carlos {Palenzuela}, and Patrick~M.
  {Motl}.
\newblock {m =1 instability and gravitational wave signal in binary neutron
  star mergers}.
\newblock {\em \prd}, 94(4):043003, August 2016.

\end{thebibliography}

\end{document}